\documentclass[12pt]{article} 
\usepackage{fullpage}
\usepackage{setspace}
\usepackage{ctable}
\usepackage{latexsym}
\usepackage{amsmath}
\usepackage{amssymb}
\usepackage{epsfig}
\usepackage{latexsym}
\usepackage{setspace}
\usepackage{multirow}
\usepackage{hyperref}
\usepackage{relsize}
\usepackage{caption}
\usepackage{subcaption}
\usepackage{natbib}
\usepackage[mdyyyy]{datetime}
\def\E{{\mathbb E}}        
\def\var{{\text{Var}}}

\def\cov{{\text{Cov}}}

\let\proglang=\textsf

\newcommand{\nn}{\nonumber\\}
\newtheorem{theorem}{Theorem}
\newtheorem{lemma}[theorem]{Lemma}

\allowdisplaybreaks

\providecommand{\keywords}[1]
{
  \small	
  \textbf{\textit{Keywords---}} #1
}

\usepackage[scaled=0.92]{helvet}	
\author{Yeng Xiong$^*$\\Michael J.~Higgins\footnote{Department of Statistics, Kansas State University. 
    We are grateful to Graeme Blair, Ben Fifield, Kosuke Imai, Ben Johnson, James Lo, the Imai Research Group, and the Higgins Research Group for helpful discussions and advice.}}  
\date{\small{\vspace{.2in}
    \begin{tabular}{cl}
    First Draft: & 1/20/2014\\ 
    This Draft: & \today  \ \ (\xxivtime) \\ [2ex]
    \end{tabular}
   }
 }
\begin{document}
\title{The Benefits of Probability-Proportional-to-Size Sampling in Cluster-Randomized Experiments}
\maketitle
\begin{abstract}

  In a cluster-randomized experiment, treatment is assigned to clusters of individual units of 
  interest--households, classrooms, villages, etc.--instead of the units themselves.  The number of clusters sampled and the number of units sampled within each cluster is typically 
  restricted by a budget constraint.  
  Previous analysis of cluster randomized experiments under the Neyman-Rubin potential outcomes
  model of response have assumed a simple random sample of clusters. Estimators
  of the population average treatment effect (PATE) under this assumption are often either
  biased or not invariant to location shifts of potential outcomes.
  We demonstrate that, by sampling clusters with probability proportional to the number of units
  within a cluster, the Horvitz-Thompson estimator (HT) is invariant to location shifts and unbiasedly estimates PATE.
  We derive standard errors of HT and discuss how to estimate these standard errors.
  We also show that results hold for stratified random samples when samples are drawn proportionally to cluster size within each stratum. 
  We demonstrate the efficacy of this sampling scheme using a simulation based on data from an experiment measuring the efficacy of the National Solidarity Programme in Afghanistan. \newline \newline
  \noindent
  \keywords{cluster randomized experiment, Neyman-Rubin model, probability-proportional-to-size sampling, Horvitz-Thompson estimator}
  
\end{abstract}

\doublespacing
\section{Introduction}

Frequently in experiments, treatment is randomized across clusters, or groups, of units of interest instead of the units themselves.  These are referred to as \textit{cluster-randomized experiments} (CREs).  
Clusters of units are often formed \textit{a priori} to the design of the experiment and without researcher intervention.  
Estimation of treatment effects is more precise when treatment is randomized across units~\citep{cornfield1978randomized}; hence, logistical issues (rather than increased precision of treatment effect estimates) motivate the randomization of treatment across clusters.  
Reasons for such randomization include addressing issues with the ethicality, legality, or feasibility of randomizing
treatment across units, reducing risk of treatment contamination, and mimicking the implementation of a proposed program (e.g. an educational intervention)~\citep{donner1998some, donner2004pitfalls, hayes2009cluster}.
Common settings for cluster-randomized experiments include: 
testing an educational intervention that is implemented within classrooms~\citep{raver2009targeting};
evaluating efficacy of a health intervention that is implemented within clinics or 
medical practices~\citep{bruce2004reducing,  king2007politically,small2008randomization, imai2009essential}; 
measuring increases in compliance and turnout from mailers sent to households~\citep{gerber2000effects};
and identifying effects of interventions implemented within villages or other geographic regions~\citep{wantchekon2003clientelism, paluck2009reducing, beath2013empowering}.

To estimate and perform inference on the \textit{population average treatment effect} (PATE), 
a CRE will require at least two stages of sampling: sampling clusters 
from a larger population of clusters (e.g.~a sample of villages within a country) 
and sampling individual units from each of the sampled clusters---samples 
may be comprised of the entire sampling frame. After a sample of clusters is obtained, but before units are sampled within each cluster, 
treatment is allocated across sampled clusters. Researchers often improve the precision of treatment effect estimates 
by drawing a stratified sample of clusters and/or blocking sampled clusters before treatment assignment~\citep{gail1996design, 
lewsey2004comparing, imai2009essential, hayes2009cluster, imbens2011experimental, hansen2014clustered}.
When researchers are interested in heterogeneous treatment effects across subpopulations of interest, within-cluster samples may also be stratified (for an example, see~\citet{kerry2005reducing}).

When clusters are sampled using simple random sampling (SRS) or stratified random sampling (StRS), current estimators of the PATE have undesirable properties.
The unbiased Horvitz-Thompson (HT-SRS) estimator~\citep{horvitz1952generalization}
is not invariant to location shifts of responses, which inflates its variance.  The location-invariant difference-in-means (DIM) estimator will be biased when treatment effects are correlated with cluster \textit{sizes}---the number of units contained within each cluster~\citep{middleton2015unbiased}.  Thus, this estimator is only unbiased in special cases such as under sharp null of no unit-level treatment effect~\citep{hansen2014clustered} or when clusters are blocked or stratified exactly on cluster sizes~\citep{donner2004pitfalls,imai2009essential}.
Moreover, as we will show, when within-cluster samples are not drawn proportional to the cluster size, DIM may estimate a quantity different from the PATE.  In fact, the only current estimator of the PATE that is both unbiased and location-invariant is the Des Raj estimator (DR)~\citep{middleton2015unbiased}, which requires the introduction of an additional parameter; however, estimating this parameter will induce bias in the estimator.  

We propose an adjustment in the \textit{design} of the experiment---as 
opposed to adjusting weights of estimators after the experiment--- for 
differences in cluster sizes: to sample clusters with \textit{probability proportional to size} 
(PPS)~\citep{hansen1943theory, cochran1977sampling, lohr2010sampling}.
We show that, under this sampling scheme, the Horvitz-Thompson estimator (HT-PPS) is both unbiased and location invariant.  

The paper is organized as follows: Section~\ref{notationsec} introduces notation.
Section~\ref{currprobs} demonstrates problems with HT-SRS, DIM, and DR estimators of PATE under
SRS of clusters. Section~\ref{haslocscaleinv} demonstrates that the HT-PPS estimator is both unbiased and location-invariant
under PPS-without-replacement sampling of clusters, gives standard errors and estimates of standard errors for HT-PPS, and shows equivalence of HT-PPS and DIM (under PPS) estimators when within-cluster sample sizes are the same across clusters.  Section~\ref{strtsection} extends results to the case where 
the sample of clusters and the within-cluster sample of units are stratified.
Section~\ref{datasection} gives simulations on a data example, which shows that the HT-PPS estimator has the smallest mean squared error compared to the other estimators.  This is due to the HT-PPS estimator being as efficient as the DIM estimator and being unbiased.  It also shows that the estimated variance is conservative for the variability of HT-PPS estimator.

\section{Notation, assumptions, and preliminaries~\label{notationsec}}

We consider a finite population of $n$ units partitioned into $\ell$ clusters.  Clusters are numbered 1 through $\ell$.
Let $n_c$ denote the number of units within cluster $c$.
Suppose units are ordered in some way within each cluster;
let $(k,c)$ denote the $k^\text{th}$ unit in cluster $c$.
We now introduce sampling and treatment assignment 
notation in the order in which they are
performed in a CRE.

\subsection{Sampling clusters}
A total of $s$ clusters are sampled; we assume $s$ is fixed and
chosen by the researcher.
Let $S_c$ denote a cluster sampling indicator;
$S_{c} = 1$ if and only if
cluster $c$ is contained in the sample.  
 \begin{equation}
        S_{c} = \left\{
        \begin{array}{ll}
          1, & \text{cluster $c$ is sampled},\\
          0, &\text{otherwise.}
        \end{array}
        \right.
\end{equation}
By definition, $\sum_{c=1}^{\ell} S_c = s$.

\subsection{Treatment assignment}

   Each of the $s$ sampled clusters is assigned to either treatment or control.
    Let $T_{ct}$ denote a treatment indicator;
$T_{ct} = 1$ if and only if
cluster $c$ receives treatment $t \in \{0,1\}$.  
 \begin{equation}
        T_{ct} = \left\{
        \begin{array}{ll}
          1, & \text{cluster $c$ receives treatment $t$},\\
          0, &\text{otherwise.}
        \end{array}
        \right.
\end{equation}
We define $T_{ct} = 0$ when $S_c = 0$.
Let $\#T_t$ denote the number of clusters that receive treatment $t$.

We suppose that treatment assignment is \textit{symmetric} across sampled clusters~\citep{miratrix2013adjusting}.
That is, conditioned on the number of treated clusters $\#T_t$, 
each of the $\binom{s}{\#T_t}$ possible treatment assignments is equally likely.  Symmetric treatment assignment implies that, for any treatment $t \in \{0,1\}$ and distinct clusters $c$, $c'$: 
	\begin{align}
		\E\left (\left.T_{ct} \right| \mathbf{S}\right) &= \frac{\#T_{t}}{s},\\
		\E\left (\left.T_{ct}T_{c't} \right| \mathbf{S}\right) &= \frac{\#T_{t}(\#T_t - 1)}{s(s-1)}.
	\end{align}
where $\mathbf S = (S_1,S_2,\ldots, S_n)$ denote a random set of cluster sampling indicator variables under a sampling design.	
Complete randomization is a special case of symmetric treatment assignment.
When the sample of clusters is stratified, 
symmetric treatment assignment also requires independence of treatment assignment 
across strata, 
which is discussed in Section~\ref{strtsection}.
	

\subsection{Within-cluster sampling}
After treatment is assigned across clusters, 
a SRS of $s_c$ units is drawn within each sampled cluster $c$.
This sample is drawn independently of treatment assignment and independently
across clusters.  We assume that these sample sizes are non-random and do not depend on the set of clusters sampled.
\noindent
Let $S_{kc}$ denote unit sampling indicator;
$S_{kc} = 1$ if and only if
the $k^\text{th}$ unit in cluster $c$ is sampled.  
 \begin{equation}
        S_{kc} = \left\{
        \begin{array}{ll}
          1, & \text{unit $(k,c)$ is sampled},\\
          0, &\text{otherwise.}
        \end{array}
        \right.
\end{equation}
We define $S_{kc} = 0$ when $S_c = 0$.
By definition, $\sum_{k=1}^{n_c} S_{kc} = s_c$.  

\subsection{Model of response: Neyman-Rubin Causal Model}
Let $y_{kct}$ denote the \textit{potential outcome} of unit $(k,c)$
given treatment $t$---the value of unit $(k,c)$ we would have observed
had that unit received treatment $t$.  
Note that $y_{kct}$ is known if and only if
that unit is sampled and receives treatment $t$ (i.e., $S_{c}T_{ct}S_{kc} = 1$).
Potential outcomes are assumed to be nonrandom.  
Let $\mathbf y = (y_{kct})_{k=1, \; c = 1, \; t = 0}^{n_c\;\;\;\;\;  \ell \;\;\;\;\;\;1}$ denote the vector of potential outcomes.

Let $Y_{kc}$ denote the observed response of unit $(k,c)$ had that 
unit been sampled.
We assume responses follow the Neyman-Rubin Causal Model (NRCM)~\citep{splawa1923application, rubin1974estimating,holland1986statistics}:
\begin{align}
  Y_{kc} &= y_{kc1}T_{c1} + y_{kc0}T_{c0} 
  \nn &= y_{kc1}T_{c1} + y_{kc0}(1-T_{c1}).
  \end{align}
Inherent in this model is the \textit{stable-unit treatment value assumption} (SUTVA), which is often referred to as the \textit{no-interference assumption}; the value of $Y_{kc}$ only depends on the treatment assigned to cluster $c$ and is not affected by the treatment assignment of any other cluster $c'$.  Observe that, since each cluster receives a single treatment condition, this assumption only needs to hold across sampled clusters and does not need to hold for units within each cluster.

\subsection{Parameter of interest}

Our quantity of interest is the \textit{population average treatment effect} (PATE):   
\begin{equation}
  \delta= \delta(\mathbf y) \equiv \sum_{c=1}^{\ell}\sum_{k=1}^{n_c} \frac{y_{kc1}-y_{kc0}}{n} = \mu_1 - \mu_0,
\end{equation}
where
\begin{equation}
  \mu_{t} = \mu_{t}(\mathbf y) \equiv \sum_{c=1}^{\ell}\sum_{k=1}^{n_c} \frac{y_{kct}}{n}.
\end{equation}
denotes the \textit{population mean for treatment $t$}.  Let
\begin{equation}
	\mu_{ct} \equiv \sum_{k=1}^{n_c} \frac{y_{kct}}{n_c}
\end{equation}
denote the \textit{population mean of cluster $c$ for treatment $t$}.
We can write the population mean as:
\begin{equation}
	\mu_{t} = \sum_{c=1}^{\ell}\sum_{k=1}^{n_c} \frac{y_{kct}}{n} 
	= \sum_{c=1}^{\ell}\frac{n_c}{n}\sum_{k=1}^{n_c} \frac{y_{kct}}{n_c}
	= \sum_{c=1}^\ell\frac{n_c}{n}\mu_{ct}.
	\label{decomposepopmean}
\end{equation}
We then define
\begin{align}
    \sigma^2_{ct} &\equiv \sum_{k=1}^{n_c} \frac{(y_{kct}-\mu_{ct})^2}{n_c-1}, \\
    \sigma^2_{t,bet} &\equiv \sum_{c=1}^\ell \frac{n_c}{n}(\mu_{ct}-\mu_t)^2,
\end{align}
respectively, as the variance of potential outcomes within cluster $c$ under treatment $t$ and as the weighted across-cluster variance for treatment $t$.

\subsection{Properties of estimators}

A function of potential outcomes $f$ is \textit{monotonically increasing} if $f(\mathbf y^*) \geq f(\mathbf y)$ whenever
\begin{equation}
	y^*_{kct} \geq y_{kct},\;\;\;\text{for all}\;\; k \in\{ 1, \ldots, n_c\},\; c \in \{1, \ldots, \ell\},\; t \in \{0,1\}.
\end{equation}
A transformation of potential outcomes $\mathbf y \rightarrow \mathbf y^*$  is \textit{linear} if, for constants 
$a, b$:
\begin{equation}
	y^*_{kct} = a + b y_{kct},\;\;\;\text{for all}\;\; k \in\{ 1, \ldots, n_c\},\; c \in \{1, \ldots, \ell\},\; t \in \{0,1\}.
\end{equation}
For simplicity, we may write this as $\mathbf y^* = a + b \mathbf y$.  A \textit{location transformation} or \textit{shift} is a linear transformation in which $b = 1$.

Observe that the population mean is a monotone increasing function that is linear in potential outcomes,
\begin{equation}
	\mu_t(a + b \mathbf y) = a + b \mu_t(\mathbf y),
\end{equation}
whereas the PATE is \textit{location-invariant}---that is, the value does not change 
given a location shift of potential outcomes,
\begin{equation}
	\delta( a + \mathbf y) = \delta(\mathbf y).
\end{equation}

\subsection{Methods for estimating PATE under SRS of clusters~\label{currprobs}}

In CREs, clusters are typically sampled using SRS, and the common estimators under this sampling procedure include the Horvitz-Thompson (HT-SRS), the difference-in-means (DIM), and the Des Raj (DR) estimators.  The HT-SRS estimator weights each unit's outcome with the inverse of the probability that the unit is treated and selected.  Therefore, it is unbiased, which  recommends it as an appropriate estimator of PATE, but \citet{imai2009essential} shows that it can be criticized on two counts.  The first is that the estimator is known to have huge variability since it does not account for varying cluster size.  Larger clusters will have greater sums of responses whereas smaller clusters will have smaller sums.  The second being that it is not location-invariant.  This poses a dilemma for variance calculation since the variance will change as $a$ changes.   The HT-SRS estimator will only be location-invariant when the number of treated \textit{units} (not clusters) is equal to the number of units assigned to control, something of which researchers cannot control.

The DIM estimator is the difference between the sample means for treated and control units.  The estimator, being elegantly simple, is favored among many researchers.  Furthermore, contrary to the HT-SRS estimator, it is efficient and invariant to location shifts.  However, \citet{middleton2015unbiased} shows that it is biased in CREs.  In actuality, the DIM estimator will be unbiased only when treatment effects are not correlated with cluster sizes and when within-cluster sample sizes are proportional to cluster sizes.

\citet{middleton2015unbiased} instead advocate the Des Raj (DR) estimator, which adds a regression component on cluster size to the HT-SRS estimator.  This helps alleviate the two criticisms on HT-SRS, but unfortunately, the solution itself poses a problem.  Estimating the regression coefficient will biased the estimator.  Having an estimate of the coefficient prior to the experiment will eliminate the bias, but this is often not feasible.  \citet{aronow2013class} expands the DR estimator to allow for additional covariates, but the same issue still persists.  In Appendix~\ref{appendixsrs}, we prove the discussed shortcomings of these estimators.

\section{Estimation of PATE under PPS sampling~\label{haslocscaleinv}}

Cluster size plays an important role in efficiently estimating the PATE in CREs.   Both the DIM and DR estimators give each cluster an equal chance of being selected, regardless of cluster size, but account for it during the estimation stage.  Staying true to the design-based philosophy of the Neyman-Rubin model, we advise instead to change the cluster sampling scheme to probability-proportional-to-size sampling (PPS), which can accommodate varying cluster sizes when sampling.  Under PPS, we derive the HT estimator, which is unbiased, location-invariant, and efficient.

\subsection{PPS sampling of clusters}

To be precise, we define a PPS sample with $s$ draws as any sample in which the probability of any cluster $c$ of being sampled is $n_cs/n$.  While, generally, PPS samples can be drawn with or without replacement, we focus exclusively on PPS samples drawn without replacement (PPSWOR), where the number of unique clusters sampled are fixed.  This allows researchers to have greater control 
in designing a CRE under a budget constraint.  A PPSWOR sampling scheme requires each cluster to contain no more than $n/s$ units.

Drawing a PPSWOR sample is a deceptively unintuitive task
and quite a bit of work has been devoted to efficient and/or exact selection of PPSWOR samples~\citep{hanurav1967optimum,vijayan1968exact, sinha1973sampling,brewer1982sampling, berger2009sampling}.  Unlike SRS or sampling with replacement, PPSWOR sampling schemes are not uniquely defined solely by the property that the
marginal probability of sampling a cluster is $n_cs/n$.
Instead, for each pair of unique clusters $c, c'$ a PPSWOR sampling scheme requires
knowing the joint probability $\pi_{cc'}$ of having both of these clusters included in the sample.
To reduce variance in estimators, it is useful to choose a sampling scheme
such that 
\begin{equation}
	\pi_{cc'} \geq P(S_c = 1)P(S_{c'} = 1) = n_cn_{c'} s^2/n^2 > 0.
	\label{jointprobineq}
\end{equation} 
\citet{sunter1977list, sunter1986solutions} provides list-sequential methods for drawing an approximate PPSWOR sample of general size $n$ satisfying~\eqref{jointprobineq}.

\subsection{Horvitz-Thompson estimator under PPS sampling}
We define the \textit{Horvitz-Thompson estimator under PPS sampling (HT-PPS) for the population mean under treatment $t$} 
as:
\begin{equation}
  \hat \mu_{t,\text{HT,PPS}} = \hat \mu_{t,\text{HT,PPS}}(\mathbf y)  \equiv \sum_{c=1}^{\ell} \frac{S_cT_{ct}}{\#T_t}\sum_{k=1}^{n_c} \frac{y_{kct}S_{kc}}{s_c}.
\end{equation}
In words, this estimate is obtained by finding each cluster $c$ that receives treatment $t$, computing the average response within each of these clusters, and then taking the average of these within-cluster averages.  The \textit{HT-PPS estimator for the PATE }is the difference of the HT-PPS estimator for the population mean under treatment and under control:
\begin{equation}
	\hat \delta_{\text{HT-PPS}} = \hat \mu_{1,\textup{HT-PPS}} - \hat \mu_{0,\textup{HT-PPS}}.
\end{equation}
Note that if a mean estimator is linear in potential outcomes, then the PATE estimator consisting of the mean estimators will be location-invariant.  The HT-PPS estimator for the population mean is linear in potential outcomes, which is formally stated in the following lemma:
\begin{lemma} \label{ppslemma}
  Suppose that clusters are sampled according to PPSWOR sampling, and suppose that treatment is symmetric across clusters.  Then:
  \begin{equation}
      \hat\mu_{t,\textup{HT, PPS}} (a + \mathbf{y}) = a + \hat\mu_{t, \textup{HT,PPS}}(\mathbf{y}).
  \end{equation}
\end{lemma}
The location invariance, unbiasedness, and variance of the HT-PPS estimator for PATE is then provided in the following theorem:

\begin{theorem} \label{ppsthm}
  Suppose that clusters are sampled according to PPSWOR sampling, and suppose that treatment is symmetric across clusters.  Then:
  \begin{align}
    \hat\delta_\text{HT,PPS}(a + \mathbf{y}) ={}& \hat\delta_\text{HT,PPS}(\mathbf{y}) \\
  	\E\left(\hat \delta_{\text{HT,PPS}}\right) ={}& \delta,\\
	\var\left(\hat \delta_{\text{HT,PPS}}\right) ={}& \sum_{t=0}^1 \left[ \E\left(\frac{1}{\#T_t}\right)\sigma^2_{t,bet} + \E\left(1-\frac{1}{\#T_t} \right)\left( \sum_{c=1}^\ell \sum_{c'\neq c} \frac{\pi_{cc'}}{s(s-1)}\mu_{ct}\mu_{c't} - \mu_t^2\right)\right]
       \nn {}& + \sum_{t=0}^1 \E\left(\frac{1}{\#T_t}\right) \sum_{c=1}^\ell \frac{n_c}{n}\left(1-\frac{s_c}{n_c}\right) \frac{\sigma^2_{ct}}{s_c} 
       \nn {}& - 2 \sum_{c=1}^{\ell} \sum_{c'\neq c}  \left[\frac{\pi_{cc'}}{s(s-1)} - \frac{n_cn_{c'}}{n^2}\right] \mu_{c1}\mu_{c'0} + 2 \sum_{c=1}^\ell \frac{n_c^2}{n^2} \mu_{c1}\mu_{c0} \label{vardeltaPPS}.
\end{align}
 \end{theorem}
The standard error for the HT-PPS estimator of PATE is then the square root of eq.~\eqref{vardeltaPPS}.  A proof of the lemma and theorem is given in Appendix~\ref{appendixhtppsvar}.  

A PPS sample naturally gives larger clusters a greater probability of being selected.  Hence, the sample will be biased towards larger clusters.  However, the HT-PPS estimator takes this into consideration as weights when estimating the PATE, thereby, eliminating the bias.  Moreover, if the same number of units are sampled from each cluster (say, $s_u$), this will give each treated (controlled) unit in the population an equal probability of being sampled, which does not hold for a SRS of clusters:
\begin{align}
    P(S_cT_{ct}S_{kc}=1 | \text{PPS}) &= \frac{\#T_ts_u}{n} \\
    P(S_cT_{ct}S_{kc}=1 | \text{SRS}) &= \frac{\#T_ts_u}{\ell n_c}.
\end{align}
Under this condition, then, the HT-PPS estimator and the DIM estimator (given a PPS sample of clusters) will be the same.  

\subsection{Variance estimator for HT-PPS estimator}

Since
\begin{equation}
    \var(\hat\delta) = \var(\hat\mu_1) + \var(\hat\mu_0) - 2\cov(\hat\mu_1, \hat\mu_0),
\end{equation}
estimating each of the three components will give an  estimator for the variance of the HT-PPS estimator for the PATE.  The Sen-Yates-Grundy (SYG) variance estimator is an unbiased estimator for the first two parts involving the sampling variance of $\hat\mu_t$ \citep{lohr2010sampling}.  On the contrary, since the last term of eq.~\eqref{vardeltaPPS} requires clusters being both treated and controlled, there is no unbiased estimator for the covariance between $\hat\mu_1$ and $\hat\mu_0$.  Consequently,  the variance of the HT-PPS estimator cannot be unbiasedly estimated, but a conservative bound is instead provided: 
\begin{align} \label{hatvarppsdel}
    \widehat{\var}_C(\hat\delta_\text{HT,PPS}) ={}& \sum_{t=0}^1 \frac{1}{2} \sum_{c=1}^\ell \sum_{c' \neq c} \left[\frac{s(s-1)}{\pi_{cc'}\#T_t(\#T_t-1)}\frac{n_cn_{c'}}{n^2} - \frac{1}{\#T_t^2} \right] S_cT_{ct}S_{c'}T_{c't} \left(\hat\mu_{ct} - \hat\mu_{c't}\right)^2 
    \nn {}& - 2\sum_{c=1}^\ell \sum_{c\neq c'} \left[ \frac{\pi_{cc'}}{s(s-1)} - \frac{n_cn_{c'}}{n^2} \right] \frac{s(s-1)}{\pi_{cc'}} \frac{S_c S_{c'} T_{c1}T_{c'0}}{\#T_1\#T_0} \hat\mu_{c1} \hat\mu_{c0}  
    \nn {}& +  \sum_{t=0}^1 \sum_{c=1}^\ell  \frac{S_cT_{ct}}{\#T_t} \frac{n_c^2}{n^2} \hat\mu_{ct}^2 
\end{align}
where $\hat\mu_{ct} = \sum_{k=1}^{n_c}y_{kct}S_{kc}/s_c$ is the within-cluster sample mean for $t$.  Note that the first term is SYG variance estimator for $\mu_t$, and the last two terms make up the covariance bound.  In appendix~\ref{appendixcovbound}, eq.~\eqref{hatvarppsdel} is shown to be positively biased for eq.~\eqref{vardeltaPPS}.  Taking the square root of eq.~\eqref{hatvarppsdel} will give the estimated standard error of the PATE estimator.

Estimating the variance requires knowledge about the $\pi_{cc'}$ under the specific sampling procedure used to obtain a PPS of clusters, but this is rarely given in practice.  Therefore, the $\pi_{cc'}$ needs to be estimated too.  This can be achieved using analytical approximations \citep{lohr2010sampling, berger2009sampling} or Monte Carlo simulations \citep{fattorini2009adaptive}.    

\section{Allowing for stratification} \label{strtsection}

Current literature recommends stratifying and/or blocking on cluster size to further reduce sampling variability.  Since PPS sampling already incorporate size variation, stratifying on other prognostic cluster covariates, rather than cluster size, can drastically improve estimation.  For example, villages may be stratified based on whether they are in a rural/urban environment or based on the villages' geographic region. Suppose that the population of $\ell$ clusters are partitioned into $m$ strata based on a categorical cluster characteristic (or a discretized numerical one).  Cluster sampling and treatment assignment is done within each stratum and independently across strata.  The cluster-stratified HT-PPS estimator is then defined (without the use of indicator variables) as 
\begin{align} \label{htppsstrtclus}
    \hat\delta_\text{CS,HT,PPS} &= \sum_{u=1}^m \frac{n_u}{n} \left[\sum^{\#T_1}_{\substack{c \in u, \\ t=1}} \frac{1}{\#T_1} \sum_{k=1}^{s_c} \frac{y_{kcu1}}{s_c} - \sum^{\#T_0}_{\substack{c' \in u, \\ t=0}} \frac{1}{\#T_0} \sum_{k^*=1}^{s_{c'}} \frac{y_{k^*c'u0}}{s_{c'}}  \right] \\
    &= \sum_{u=1}^m \frac{n_u}{n} \hat\delta_{u, \text{HT,PPS}}
\end{align}
where $n_u$ is the population of units in stratum $u$.  The statistical properties for the cluster-stratified HT-PPS estimator can be easily derived from Theorem \ref{ppsthm}.
\begin{theorem}
  Suppose clusters are first stratified.  Suppose also that clusters are sampled with PPSWOR and treatments are randomized within stratum and independently across strata.  Then:
  \begin{align}
      \E\left(\hat\delta_\textup{CS,HT,PPS}\right) &= \delta \\
      \var\left(\hat\delta_\textup{CS,HT,PPS}\right) &= \sum_{u=1}^m \frac{n_u^2}{n^2} \var\left(\hat\delta_{u, \textup{HT,PPS}}\right) \label{ppsstrtvar} \\
      \hat\delta_\textup{CS,HT,PPS}(a + \mathbf{y}) &= \hat\delta_\textup{CS,HT,PPS}(\mathbf{y}). 
  \end{align}
\end{theorem}  
Plugging eq.~\eqref{hatvarppsdel} into eq.~\eqref{ppsstrtvar} will give a conservative estimate of the sampling variability for the cluster-stratified HT-PPS estimator.  

Stratification may be applied to units within clusters instead of on clusters.  In this setting, the $n_c$ units in cluster $c$ are divided into $q_c$ strata with $n_{v}$ units in each stratum.  A SRS sample of $s_{v}$ units is taken.  The unit-stratified HT-PPS estimator is
\begin{equation} \label{htppsstrtunit}
    \hat\delta_\text{US,HT,PPS} = \sum^{\#T_1}_{\substack{c=1,\\ t=1}} \frac{1}{\#T_1} \sum^{q_c}_{v \in c} \frac{n_{v}}{n_c} \sum_{k \in v}^{s_{v}} \frac{y_{kvc1}}{s_{v}} - \sum^{\#T_0}_{\substack{c=1,\\t=0}} \frac{1}{\#T_0} \sum_{v \in c}^{q_c} \frac{n_{v}}{n_c} \sum_{k \in v}^{s_{v}} \frac{y_{kvc0}}{s_{v}}.
\end{equation}
Since the stratification is on the units within a cluster, we need to only adjust the within-cluster variance in Theorem \ref{ppsthm} to get the statistical properties for the unit-stratified HT-PPS estimator.  Hence,
\begin{equation}
    \var(\hat\mu_{ct}) = \frac{1}{\#T_t} \sum_{c=1}^\ell \frac{n_c}{n}\left(1-\frac{s_c}{n_c}\right) \frac{\sigma^2_{ct}}{s_c}
\end{equation}
will instead be
\begin{equation}
    \var(\hat\mu_{ct}) = \frac{1}{\#T_t} \sum_{c=1}^\ell \frac{n_c}{n} \sum_{v\in c}^{q_c} \frac{n_v^2}{n_c^2} \left(1-\frac{s_v}{n_v}\right)\frac{\sigma^2_{ct}}{s_v}.
\end{equation}
Similarly, eq.~\eqref{hatvarppsdel} is still a conservative estimate of the sampling variability, but $\hat\mu_{ct}$ will instead be the stratified estimator of the within-cluster sample mean for $t$. Naturally, if stratification is desired at both the cluster- and unit-levels, combining eq.~\eqref{htppsstrtclus} and eq.~\eqref{htppsstrtunit} will give an unbiased estimator for the PATE. 

\section{Data example} \label{datasection}

\citet{beath2013empowering} perform an experiment in Afghanistan to investigate whether development programs with mandatory women contribution can change villagers' perspectives on women's political participation.  Five hundred villages, ranging from sizes 60 to 9000, were sampled and matched into pairs.  Within each pair, one village is randomly assigned to receive the National Solidarity Programme (NSP), and the other village serves as a control to receive the NSP after the experiment.  The NSP creates a community development council and provides grants for village development projects.  The council is then responsible for distributing the grants among the projects.  However, the NSP stipulates that half of the council must be women and at least one of the projects must be a priority for the women. After two years, ten head-of-household men and ten head-of-household women from each village are selected for a follow-up survey.  Respondents are asked whether women should have equal decision making in the village council.  

\begin{figure}
\centering
\begin{subfigure}{\textwidth}
  \centering
  \includegraphics[width=0.85\linewidth]{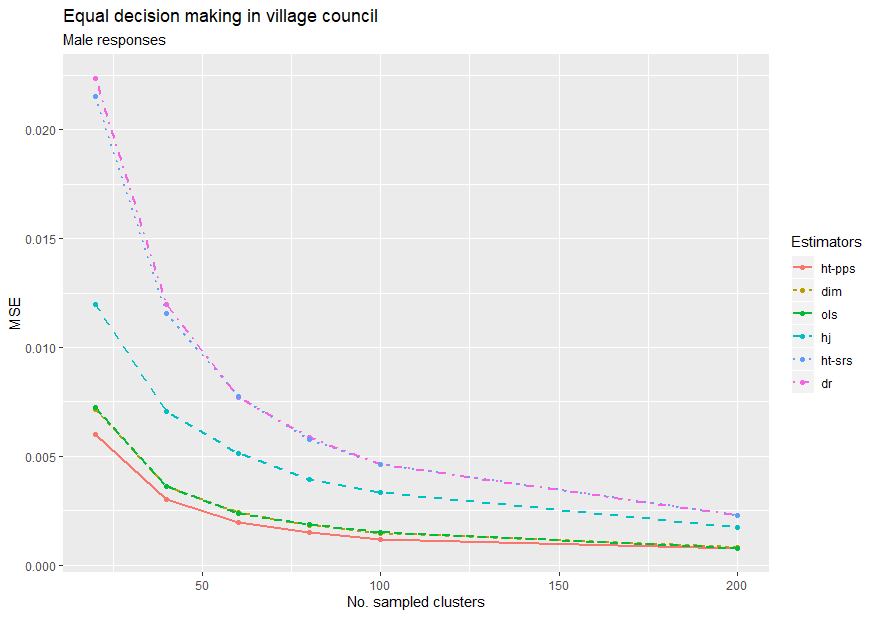}
  \label{fig:sub1}
\end{subfigure} \newline
\begin{subfigure}{\textwidth}
  \centering
  \includegraphics[width=0.85\linewidth]{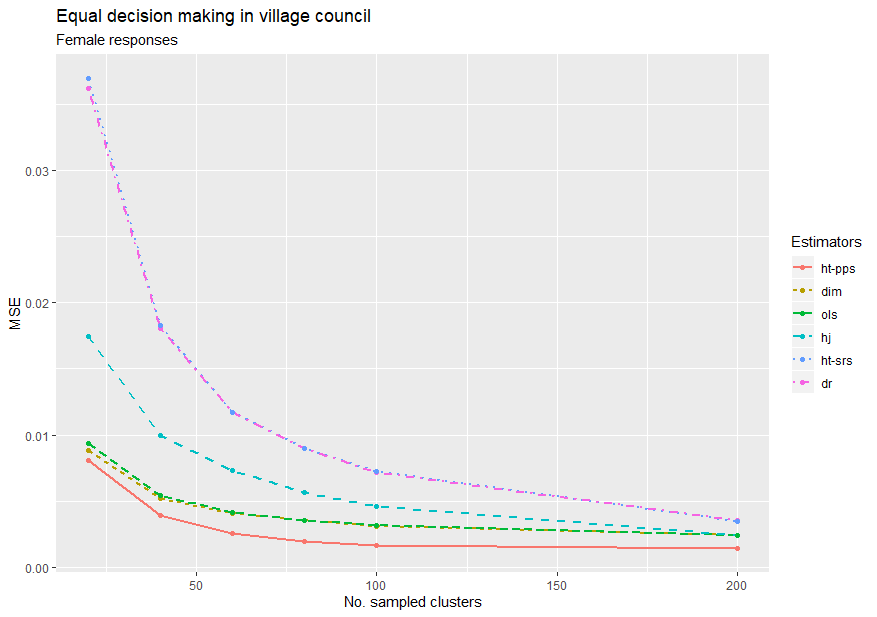}
  \label{fig:sub2}
\end{subfigure}
\caption{The number of sampled clusters are 20, 40, 60, 80, 100, and 200.  Results are based on 10,000 simulations.  Our estimator, HT-PPS (red and solid), performs better than the SRS estimators.}
\label{fig:mse}
\end{figure}

We perform Monte Carlo simulations to compare the HT-PPS estimator to its SRS counterparts.  We generate the potential outcomes from a fitted LOWESS line of the NSP data and then mimic a simplified experiment in which clusters are randomly sampled using either PPS or SRS. The \proglang{R} package \textit{TeachingSampling} is used to perform Sunter's PPSWOR sampling.  We vary the number of sampled clusters from 20 to 200.  Treatments are assigned completely at random to the sampled clusters.  For ease, we fix the number of treated clusters to be half of the sampled clusters, but this will not drastically change the theoretical results.

The PATE is then estimated with the HT-PPS, DIM, HT-SRS, biased DR, and H\'{a}jek estimators.  For the DR estimator, $\theta$ is optimally estimated as described in \citet{middleton2015unbiased} using the simulated sample data.  The H\'{a}jek estimator (see \citep{hajek1971discussion}) is a ratio estimator similar to the DIM.  It estimates the population mean for treatment $t$ as a ratio of the estimated treated (controlled) cluster total over the total number of treated (controlled) units in the sampled clusters.  The other estimators are as described in section \ref{currprobs}.  

\begin{figure}
\centering
\begin{subfigure}{\textwidth}
  \centering
  \includegraphics[width=0.85\linewidth]{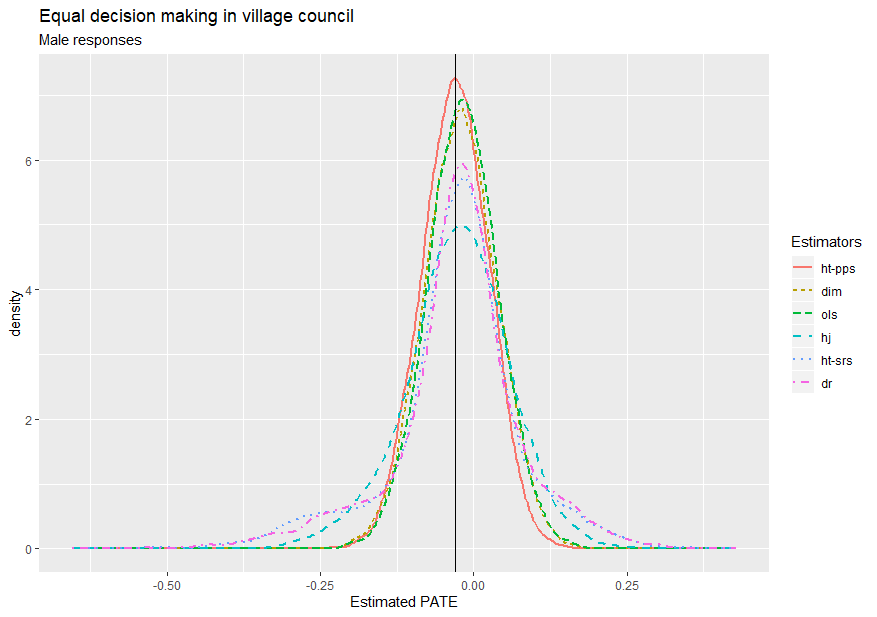}
\end{subfigure} \newline
\begin{subfigure}{\textwidth}
  \centering
  \includegraphics[width=0.85\linewidth]{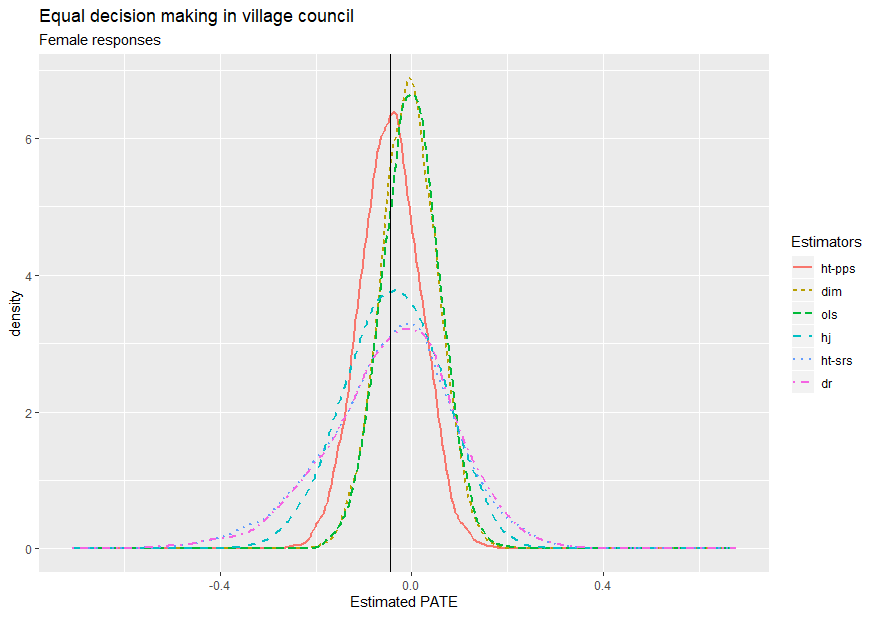}
\end{subfigure}
\caption{Results based on 10,000 simulations of sampling 40 clusters.  The solid vertical line is the PATE (-0.0302 for male, -0.0448 for female).  Our estimator, HT-PPS (red and solid), is unbiased and as efficient as the DIM.}
\label{fig:density}
\end{figure}

Figure~\ref{fig:mse} compares the MSE of the estimators as the number of sampled clusters are varied.  To get an exact PPSWOR sample of the NSP data, the number of sampled clusters can be at most 45, and thus, the samples of sizes 60 and up are only approximately PPS.  Even so, the HT-PPS estimator performs best out of all the estimators, including the omnipresent DIM, across all sample sizes of clusters.  Figure~\ref{fig:density} gives a more thorough comparison of the sampling distributions for the PATE estimators when 40 clusters are sampled.

In addition, Monte Carlo simulations are done to examine the performance of estimating the sampling variance of the HT-PPS estimator.  The $\pi_{cc'}$ are estimated using analytical approximations.  Table~\ref{table} provides statistics (estimated variance, bias, and true sampling variability) on the variance estimation as the number of sampled clusters are varied.  Note that the bias is all positive so estimates are conservative.  Estimates are also close to the true variance.   

\begin{table}[ht]
\centering
\begin{tabular}{ |c|c|c|c||c|c|c| } 
 \hline
 & \multicolumn{3}{c||}{Male responses} & \multicolumn{3}{c|}{Female responses} \\ \hline
 No. sampled clusters & Est.~var. & Bias & Samp.~var. & Est.~var. & Bias & Samp.~var.\\ \hline
 20 & 0.0061 & 6.47E-05 & 6.37E-06 & 0.0082 & 2.48E-04 & 4.76E-06 \\
 40 & 0.0031 & 7.15E-05 & 7.39E-07 & 0.0041 & 1.9E-04 & 5.23E-07 \\
 60\footnotemark & 0.0034 & 1.45E-03 & 4.83E-07 & 0.0069 & 4.29E-03 & 6.68E-07 \\
 80\footnotemark[\value{footnote}] & 0.0026 & 1.13E-03 & 2.06E-07 & 0.0055 & 3.56E-03 & 2.84E-07 \\
 100\footnotemark[\value{footnote}] & 0.0021 & 9.44E-04 & 1.07E-07 & 0.0046 & 3.07E-03 & 1.47E-07 \\
 200\footnotemark[\value{footnote}] & 0.0012 & 6.08E-04 & 1.31E-08 & 0.0026 & 1.92E-03 & 1.79E-08  \\
 \hline
\end{tabular}
\caption{Results based on 10,000 simulations.  Estimated variances are upwardly biased, but the bias is marginally small.}
\label{table}
\end{table}
\footnotetext{Estimated variances uses the with-replacement variance estimator since samples are not exactly PPS.}

\section{Conclusion}

Experiments are the ``gold standard'' for investigating causal relationships, but traditionally, the causal inference is limited to the convenience sample recruited for the experiment.  Often, though, researchers prefer to generalize to individuals beyond those in the sample.  This then requires a random sample from the population of interest.  Since populations are naturally structured in groups, it is easier to sampled groups, rather than individuals, to be in experiments; thus, cluster randomized experiments are a fitting design choice.

On the other hand, the multi-level constitution of CREs poses analytical adversities.  Much of the difficulties arises from unequal cluster sizes.   If clusters contain the same number of units, all estimators discussed would be the same, and the idea of choosing the ``best'' estimator would be nonexistent.  However, varying cluster sizes are intrinsic to CREs.  Hence, in this paper, we account for cluster sizes by sampling clusters with probability proportional to size.  Estimating PATE can then be done with the HT-PPS estimator.  The HT-PPS estimator is an attractive alternative to SRS-based estimators since it is intuitive, unbiased, efficient, and location-invariant.  We also derive a conservative variance estimator for the sampling variability of the HT-PPS estimator.

Stratification and blocking can still be used to further reduce the sampling variability, but with PPS sampling, other more important covariates can be used instead of cluster size. We have done some work on how stratification may affect the HT-PPS estimator, but we plan to expand on it.  We also plan on extending our results to include blocking too.     

\singlespacing
  \bibliography{CREref.bib}  
  \doublespacing
\newpage
\appendix
\cleardoublepage

\section{Properties of the SRS estimators} \label{appendixsrs}

\subsection{Horvitz-Thompson estimator}

The \textit{Horvitz-Thompson (HT-SRS) estimator for the population mean under treatment $t$} is defined as:
\begin{equation} \label{htmu}
  \hat \mu_{t,\text{HT,SRS}} =\ell\sum_{c=1}^{\ell} \frac{S_cT_{ct}}{\#T_t}\frac{n_c}{n}\sum_{k=1}^{n_c} \frac{y_{kct}S_{kc}}{s_c}.
\end{equation}
\textit{The HT-SRS estimator for the PATE} is then the difference of the HT-SRS estimator for the population mean under treatment and under control:
\begin{equation}
	\hat \delta_{\text{HT, SRS}} = \hat \mu_{1,\text{HT,SRS}}  - \hat \mu_{0,\text{HT,SRS}} .
\end{equation}
The HT-SRS estimator for PATE is not location-invariant since
\begin{align}
    \hat\delta_\text{HT,SRS} (a + \mathbf{y}) ={}& \hat \mu_{1,\text{HT,SRS}}(a + \mathbf y) - \hat \mu_{0,\text{HT,SRS}}(a + \mathbf y)
    \nn ={}& \ell\sum_{c=1}^{\ell} \frac{S_cT_{c1}}{\#T_1}\frac{n_c}{n}\sum_{k=1}^{n_c} \frac{(a+y_{kc1})S_{kc}}{s_c} - \ell\sum_{c=1}^{\ell} \frac{S_cT_{c0}}{\#T_0}\frac{n_c}{n}\sum_{k=1}^{n_c} \frac{(a+y_{kc0})S_{kc}}{s_c}
	\nn ={}& a \left(\ell\sum_{c=1}^{\ell} \frac{S_cT_{c1}}{\#T_1}\frac{n_c}{n}\sum_{k=1}^{n_c} \frac{S_{kc}}{s_c} - \ell\sum_{c=1}^{\ell} \frac{S_cT_{c0}}{\#T_0}\frac{n_c}{n}\sum_{k=1}^{n_c} \frac{S_{kc}}{s_c} \right) 
	\nn {}& + \left( \ell\sum_{c=1}^{\ell} \frac{S_cT_{c1}}{\#T_1}\frac{n_c}{n}\sum_{k=1}^{n_c} \frac{y_{kc1}S_{kc}}{s_c} - \ell\sum_{c=1}^{\ell} \frac{S_cT_{c0}}{\#T_0}\frac{n_c}{n}\sum_{k=1}^{n_c} \frac{y_{kc0}S_{kc}}{s_c} \right)
	\nn={}& a \left( \frac{\ell\#N_1}{n \#T_1} -\frac{\ell\#N_0}{n \#T_0} \right) +  \hat\delta_\text{HT,SRS}(\mathbf{y}),
\end{align}
where
\begin{equation}
    \#N_t = \sum_{c=1}^\ell S_cT_{ct}n_c
\end{equation}
represents all units given treatment $t$ in the sampled clusters.   Lack of location-invariance will not affect the unbiasedness of the estimator, even for linearly transformed outcomes, since 
\begin{equation}
    \E(\#N_t) = \sum_{c=1}^\ell \E(S_cT_{ct}n_c) = \frac{n\#T_t}{\ell}.
\end{equation}
However, this problem presents itself in the variance calculation: \begin{align}
   	\var\left(\hat{\delta}_{\text{HT,SRS}}(a+\mathbf{y})\right) ={}& a^2 \left(\frac{\ell}{n}\right)^2 \left[\var\left(\frac{\#N_1}{\#T_1}\right) + \var\left(\frac{\#N_0}{\#T_0}\right)-2\cov\left(\frac{\#N_1}{\#T_1}, \frac{\#N_0}{\#T_0}\right)\right] 
    \nn {}& + 2a\frac{\ell}{n} \left[ \cov\left(\frac{\#N_1}{\#T_1}, \hat{\delta}_{\text{HT,SRS}}\right) - \cov\left(\frac{\#N_0}{\#T_0}, \hat{\delta}_{\text{HT,SRS}}\right) \right] 
    \nn {}& + \var(\hat{\delta}_{\text{HT,SRS}}).
\end{align}

\subsection{Difference-in-means estimator}

The \textit{sample mean under treatment $t$}
is:
\begin{equation}
  \hat \mu_{t,\text{DIM,SRS}} \equiv \frac{\sum_{c=1}^{\ell} S_cT_{ct} \sum_{k=1}^{n_c} y_{kct}S_{kc}}{\sum_{c=1}^{\ell}S_cT_{ct}s_c}. 
\end{equation}
\textit{The difference-in-means (DIM) estimator for the PATE} is the difference of the sample means under treatment and under control:
\begin{equation}
	\hat \delta_{\text{DIM, SRS}} = \hat \mu_{1,\text{DIM,SRS}}  - \hat \mu_{0,\text{DIM,SRS}}.
\end{equation}
Using the relationship
\begin{align}
    \E\left(\frac{u}{v}\right) = \frac{1}{\E(v)}\left[\E(u) - \cov(\frac{u}{v}, v)\right],
\end{align}
it can be shown that the DIM estimator is actually estimating the quantity
\begin{align}
    \E(\hat\mu_{t,\text{DIM, SRS}}) = \frac{1}{\sum_{c=1}^\ell s_c}\left[ \sum_{c=1}^\ell \sum_{k=1}^{n_c} \frac{s_c}{n_c} y_{kct} - \ell\cov\left( \hat\mu_{t,\text{DIM, SRS}}, \sum_{c=1}^\ell \frac{S_cT_{ct}s_c}{\#T_t} 
    \right)\right].
\end{align}

\subsection{Des Raj estimate of PATE under SRS}

\citet{middleton2015unbiased} advocate the Des Raj (DR) estimator, and they define the \textit{DR estimator for the population mean under treatment $t$} as:
\begin{equation}
  \hat \mu_{t,\text{DR,SRS}} = 
   \ell \sum_{c=1}^\ell \frac{S_cT_{ct}}{\#T_t}\frac{n_c}{n}\left[\sum_{k=1}^{n_c}\frac{y_{kct}S_{kc}}{s_c}-\frac{\theta}{n_c}\left(n_c-\frac{n}{\ell} \right) \right]
\end{equation}
  where $\theta$ is a regression coefficient.  \textit{The DR estimator for the PATE} is then the difference of the DR estimator for the population mean under treatment and under control:
\begin{equation}
	\hat \delta_{\text{DR, SRS}} = \hat \mu_{1,\text{DR,SRS}}  - \hat \mu_{0,\text{DR,SRS}} .
\end{equation}
We show here that the DR estimator is biased when the regression coefficient $\theta$ needs to be estimated:
\begin{align}
    \E(\hat\delta_\text{DR,SRS}) ={}& \delta - \ell \sum_{c=1}^\ell \frac{n_c}{n} \cov\left(\frac{S_cT_{c1}}{\#T_1}, \hat\theta\right) 
    \nn {}& + \ell \sum_{c=1}^\ell 
    \frac{n_c}{n} \cov\left(\frac{S_cT_{c0}}{\#T_0}, \hat\theta\right).
\end{align}
\cleardoublepage

\section{Properties of HT-PPS estimator} \label{appendixhtppsvar}

We prove the results for Lemma~\ref{ppslemma} and Theorem~\ref{ppsthm}.  For more detailed derivations, we provide an expanded supplemental appendix.

\subsection{Linearity in potential outcomes for HT-PPS mean estimator}

\begin{align}
    \hat\mu_{t,\text{HT,PPS}}(a + \mathbf{y}) &= \sum_{c=1}^{\ell}  \frac{S_cT_{ct}}{\#T_t}\sum_{k=1}^{n_c} \frac{(a + y_{kct})S_{kc}}{s_c}
    \nn &= 	a\sum_{c=1}^{\ell}  \frac{S_cT_{c1}}{\#T_1}\sum_{k=1}^{n_c} \frac{S_{kc}}{s_c} + \sum_{c=1}^{\ell}  \frac{S_cT_{c1}}{\#T_1}\sum_{k=1}^{n_c} \frac{y_{kc1}S_{kc}}{s_c}
    \nn &= a + \hat\mu_{t,\text{HT,PPS}}(\mathbf{y})
\end{align}



\subsection{Expectation of HT-PPS estimator for PATE}

  \begin{align}
  \E(\hat\delta_{\text{HT-PPS}}) ={}& \E\left(\sum_{c=1}^{\ell}  \frac{S_cT_{c1}}{\#T_1}\sum_{k=1}^{n_c} \frac{y_{kc1}S_{kc}}{s_c} - \sum_{c=1}^{\ell}  \frac{S_{c'}T_{c'0}}{\#T_0}\sum_{k^*=1}^{n_{c'}} \frac{y_{k^*c'0}S_{k^*c'}}{s_{c'}}\right) 
    \nn ={}& \sum_{c=1}^{\ell} \sum_{k=1}^{n_c} \frac{y_{kc1}}{ s_c}\E\left( S_c\E\left(\left.\frac{T_{c1}}{\#T_1}\right|\mathbf{S}\right)\E(S_{kc}|\mathbf{S})  \right) 
    \nn {}& - \sum_{c'=1}^{\ell} \sum_{k^*=1}^{n_{c'}} \frac{y_{k^*c'0}}{ s_{c'}}\E\left( S_{c'}\E\left(\left.\frac{T_{c'0}}{\#T_0}\right|\mathbf{S}\right)\E(S_{k^*c'}|\mathbf{S})  \right)
    \nn ={}& \sum_{c=1}^{\ell} \sum_{k=1}^{n_c} \frac{y_{kc1}}{ n_c s}\E\left( S_c\right) - \sum_{c'=1}^{\ell} \sum_{k^*=1}^{n_{c'}} \frac{y_{k^*c'0}}{ n_{c'}s}\E\left( S_{c'}\right)    
    \nn ={}& \sum_{c=1}^\ell \sum_{k=1}^{n_c} \frac{y_{kc1}}{n} - \sum_{c'=1}^\ell \sum_{k^*=1}^{n_{c'}} \frac{y_{k^*c'0}}{n}
    \nn ={}& \mu_1-\mu_0=\delta.
  \end{align}

\subsection{Variance of HT-PPS estimator for PATE}

From the property
\begin{equation}
    \var(\hat\delta) = \var(\hat\mu_1 - \hat\mu_0) = \var(\hat\mu_1) + \var(\hat\mu_0) - 2\cov(\hat\mu_1, \hat\mu_0).
\end{equation}
each term is expanded upon to derive the variance of the HT-PPS estimator for PATE and obtain a variance estimator.

\subsubsection{Variance of HT-PPS estimator for population mean} \label{varhtpps}

Using the law of total variance,
   \begin{align} 
       \var(\hat\mu_{t,\text{HT,PPS}}) ={}& \var\left(\sum_{c=1}^\ell \frac{S_cT_{ct}}{\#T_t} \sum_{k=1}^{n_c}\frac{y_{kct}S_{kc}}{s_c} \right)
       \nn ={}& \var\left[ \E\left(\left.\sum_{c=1}^\ell \frac{S_cT_{ct}}{\#T_t}\sum_{k=1}^{n_c}\frac{y_{kct}S_{kc}}{s_c} \right| \mathbf{S, T} \right) \right] 
       \nn {}& + \E\left[\var\left(\left.\sum_{c=1}^\ell \frac{S_cT_{ct}}{\#T_t}\sum_{k=1}^{n_c}\frac{y_{kct}S_{kc}}{s_c} \right| \mathbf{S, T} \right) \right].
\end{align}
The first terms can be further simplified:
\begin{align}
       {}&  \var\left[ \E\left(\left.\sum_{c=1}^\ell \frac{S_cT_{ct}}{\#T_t}\sum_{k=1}^{n_c}\frac{y_{kct}S_{kc}}{s_c} \right| \mathbf{S, T} \right) \right] 
       \nn {}& = \sum_{c=1}^\ell \mu_{ct}^2 \var\left(\frac{S_cT_{ct}}{\#T_t}\right)  + \sum_{c=1}^\ell \sum_{c'\neq c} \mu_{ct}\mu_{c't}\cov\left(\frac{S_cT_{ct}}{\#T_t}, \frac{S_{c'}T_{c't}}{\#T_t}\right)
       \nn {}&= \sum_{c=1}^\ell \mu_{ct}^2 \left[\E\left(\frac{S_cT_{ct}}{\#T_t}\right) - \E\left(\frac{S_cT_{ct}}{\#T_t}\right)^2 \right] 
       \nn {}& \hspace{3ex} + \sum_{c=1}^\ell \sum_{c'\neq c} \mu_{ct}\mu_{c't} \left[\E\left(\frac{S_cS_{c'}T_{ct}T_{c't}}{\#T_t^2}\right)-\E\left(\frac{S_cT_{ct}}{\#T_t}\right)\E\left(\frac{S_{c'}T_{c't}}{\#T_t}\right) \right]
       \nn {}& = \E\left(\frac{1}{\#T_t}\right) \sum_{c=1}^\ell \frac{n_c}{n} \mu_{ct}^2 -\sum_{c=1}^\ell \sum_{c'\neq c} \frac{n_c^2}{n^2}\mu_{ct}^2 
       \nn {}& \hspace{3ex} + \E\left(1-\frac{1}{\#T_t}\right)\sum_{c=1}^\ell \sum_{c'\neq c}\frac{\pi_{cc'}}{s(s-1)} \mu_{ct}\mu_{c't} - \sum_{c=1}^\ell \sum_{c'\neq c}\frac{n_c^2}{n^2}\mu_{ct}\mu_{c't}
       \nn {}& = \E\left(\frac{1}{\#T_t}\right)\sigma^2_{t,bet} + \E\left(1-\frac{1}{\#T_t}\right) \left[\sum_{c=1}^\ell \sum_{c'\neq c}\frac{\pi_{cc'}}{s(s-1)}\mu_{ct}\mu_{c't} - \mu_t^2 \right]
\end{align}   
where $\sigma^2_{t,bet}$ is the weighted variance of cluster means.  Simplifying the second term:
\begin{align}       
       & \E\left[\var\left(\left.\sum_{c=1}^\ell \frac{S_cT_{ct}}{\#T_t}\sum_{k=1}^{n_c}\frac{y_{kct}S_{kc}}{s_c} \right| \mathbf{S, T} \right) \right] =  \sum_{c=1}^\ell \var(\hat\mu_{ct}|\mathbf{S, T}) \E\left(\frac{S_cT_{ct}}{\#T_t^2}\right)
       \nn {}& = \E\left(\frac{1}{\#T_t}\right) \sum_{c=1}^\ell \frac{n_c}{n}\var(\hat\mu_{ct})
       \nn {}& = \E\left( \frac{1}{\#T_t}\right) \sum_{c=1}^\ell \frac{n_c}{n}\left(1-\frac{s_c}{n_c}\right) \frac{\sigma^2_{ct}}{s_c}.
       \end{align}
The variance for the HT-PPS mean estimator is
\begin{align}
    \var(\hat\mu_{t, \text{HT, SRS}}) ={}& \E\left(\frac{1}{\#T_t}\right) \sum_{c=1}^\ell \frac{n_c}{n}\mu_{ct}^2  + \E\left(1-\frac{1}{\#T_t}\right)\sum_{c=1}^\ell \sum_{c\neq c'} \frac{\pi_{cc'}}{s(s-1)}  \mu_{ct}\mu_{c't} - \mu_t^2 
    \nn {}& + \E\left(\frac{1}{\#T_t}\right)\sum_{c=1}^\ell \frac{n_c}{n}\var(\hat\mu_{ct}) \label{varpps1} \\
     ={}& \E\left(\frac{1}{\#T_t}\right)\sigma^2_{t,bet} + \E\left(1-\frac{1}{\#T_t}\right) \left[\sum_{c=1}^\ell \sum_{c'\neq c}\frac{\pi_{cc'}}{s(s-1)}\mu_{ct}\mu_{c't} - \mu_t^2 \right] 
    \nn {}& + \E\left( \frac{1}{\#T_t}\right) \sum_{c=1}^\ell \frac{n_c}{n}\left(1-\frac{s_c}{n_c}\right) \frac{\sigma^2_{ct}}{s_c}.
\end{align}

\subsubsection{SYG estimator for variance}

The SYG variance estimator is
\begin{align}
    \widehat\var(\hat\mu_t) ={}& \frac{1}{2} \sum_{c=1}^\ell \sum_{c\neq c'} \left[\frac{s(s-1)}{\pi_{cc'}\#T_t(\#T_t-1)}\frac{n_cn_{c'}}{n^2} - \frac{1}{\#T_t^2} \right] S_cT_{ct}S_{c'}T_{c't} (\hat\mu_{ct}-\hat\mu_{c't})^2 
    \nn {}& +  \sum_{c=1}^\ell \frac{S_cT_{ct}}{\#T_t} \frac{n_c}{n} \widehat{\var}(\hat\mu_{ct})
\end{align}
where 
\begin{equation}
    \widehat{\var}(\hat\mu_{ct}) = \left(1-\frac{s_c}{n_c}\right) \frac{\hat\sigma^2_{ct}}{s_c}.
\end{equation}
The $\hat\sigma^2_{ct}$ is the sample variance of outcomes, which is unbiased for the population variance $\sigma^2_{ct}$.  We will now show that the SYG variance is unbiased for $\var(\hat\mu_t)$.  
\begin{align}
    {}& \E\left(\widehat\var(\hat\mu_t)\right) 
    \nn {}& = \E\left( \E\left(\left.\frac{1}{2} \sum_{c=1}^\ell \sum_{c\neq c'} \left[ \frac{s(s-1)}{\pi_{cc'}} \frac{n_cn_{c'}}{n^2} \frac{S_cS_{c'}T_{ct}T_{c't}}{\#T_t(\#T_t-1)} - \frac{S_cS_{c'}T_{ct}T_{c't}}{\#T_t^2} \right]  [\hat\mu_{ct}-\hat\mu_{c't}]^2\right|\mathbf{S, T}\right)\right)
    \nn {}& \hspace{3ex} + \E\left(\E\left(\left. \sum_{c=1}^\ell \frac{n_c}{n} \frac{S_cT_{ct}}{\#T_t}\widehat{\var}(\hat\mu_{ct})\right|\mathbf{S,T}\right)\right)
    \nn ={}&  \sum_{c=1}^\ell \sum_{c\neq c'} \left[ \frac{s(s-1)}{\pi_{cc'}} \frac{n_cn_{c'}}{n^2} \E\left(\frac{S_cS_{c'}T_{ct}T_{c't}}{\#T_t(\#T_t-1)}\right) - \E\left(\frac{S_cS_{c'}T_{ct}T_{c't}}{\#T_t^2}\right) \right]  [\mu^2_{ct}+\var(\hat\mu_{ct})-\mu_{ct}\mu_{c't}]
    \nn {}& +  \sum_{c=1}^\ell \frac{n_c}{n} \var(\hat\mu_{ct}) \E\left(\frac{S_cT_{ct}}{\#T_t}\right) 
    \nn ={}&  \sum_{c=1}^\ell \sum_{c\neq c'} \left[ \frac{s(s-1)}{\pi_{cc'}} \frac{n_cn_{c'}}{n^2} \E\left(\frac{\E\left(\left.S_cS_{c'}T_{ct}T_{c't} \right| \#T_t \right)}{\#T_t(\#T_t-1)}\right) - \E\left( \frac{\E\left(\left.S_cS_{c'}T_{ct}T_{c't} \right| \#T_t\right)}{\#T_t^2}\right) \right]  
    \nn {}& \hspace{3ex} \cdot  [\mu^2_{ct}+\var(\hat\mu_{ct})-\mu_{ct}\mu_{c't}] +  \sum_{c=1}^\ell \frac{n_c}{n} \var(\hat\mu_{ct}) \E\left(\frac{\E\left(S_cT_{ct}|\#T_t\right)}{\#T_t}  \right)
    \nn ={}& \E\left(\frac{1}{\#T_t}\right) \sum_{c=1}^\ell \frac{n_c}{n}\mu_{ct}^2  + \E\left(1-\frac{1}{\#T_t}\right)\sum_{c=1}^\ell \sum_{c\neq c'} \frac{\pi_{cc'}}{s(s-1)}  \mu_{ct}\mu_{c't} - \mu_t^2 
    \nn {}& + \E\left(\frac{1}{\#T_t}\right)\sum_{c=1}^\ell \frac{n_c}{n}\var(\hat\mu_{ct}). \label{varppsexp}
\end{align}
This is equal to eq.~\eqref{varpps1}.    

\subsubsection{Covariance of HT-PPS estimator for population means}	\label{covhtpps}
	Note that:
   \begin{align}
       \hat \mu_{1,\text{HT,PPS}}\hat \mu_{0,\text{HT,PPS}} & = \left( \sum_{c=1}^{\ell} \frac{S_cT_{c1}}{\#T_1}\sum_{k=1}^{n_c} \frac{y_{kc1}S_{kc}}{ s_c}\right) \left( \sum_{c'=1}^{\ell} \frac{S_{c'}T_{c'0}}{\#T_0}\sum_{k^*=1}^{n_{c'}} \frac{y_{k*c'0}S_{k*c'}}{ s_{c'}}\right)
        \nn &= \sum_{c=1}^{\ell} \sum_{k=1}^{n_c}\sum_{c'\neq c} \sum_{k^*=1}^{n_{c'}} \frac{y_{kc1}y_{k^*c'0}}{s_cs_{c'}} \frac{S_cT_{c1}S_{kc}S_{c'}T_{c'0}S_{k^*c'}}{\#T_1\#T_0}.
   \end{align}
Therefore,
   \begin{align}
       {}& \cov(\hat \mu_{1,\text{HT,PPS}}, \hat \mu_{0,\text{HT,PPS}})  = \E(\hat \mu_{1,\text{HT,PPS}}\hat \mu_{0,\text{HT,PPS}}) - \E(\hat \mu_{1,\text{HT,PPS}})\E(\hat \mu_{0,\text{HT,PPS}})
       \nn {}&= \sum_{c=1}^{\ell} \sum_{k=1}^{n_c}\sum_{c'\neq c} \sum_{k^*=1}^{n_{c'}} \frac{y_{kc1}y_{k^*c'0}}{s_cs_{c'}}  \E\left[S_cS_{c'} \E\left(\left.\frac{T_{c1}T_{c'0}}{\#T_1\#T_0}\right|\mathbf{S}\right) \E(S_{kc}S_{k^*c'}|\mathbf{S})\right] - \mu_1\mu_0
       \nn {}&= \sum_{c=1}^{\ell} \sum_{c'\neq c}  \left[\frac{\pi_{cc'}}{s(s-1)} - \frac{n_cn_{c'}}{n^2}\right] \mu_{c1}\mu_{c'0} - \sum_{c=1}^\ell \frac{n_c^2}{n^2} \mu_{c1}\mu_{c0}. \label{covpps}
   \end{align}
   
   \subsubsection{Covariance bound} \label{appendixcovbound}

The covariance is bounded by
\begin{align}
    \widehat{\cov}_C(\hat\mu_1, \hat\mu_0) ={}& \sum_{c=1}^\ell \sum_{c'\neq c} \left[1 -\frac{n_cn_{c'}}{n^2}\frac{s(s-1)}{\pi_{cc'}}\right]   \frac{S_cT_{ct}S_{c'}T_{c't}}{\#T_1 \#T_0} \hat\mu_{c1}\hat\mu_{c'0}
    \nn {}& - \frac{1}{2} \sum_{c=1}^\ell \frac{n_c}{n} \frac{S_cT_{c1}}{\#T_1} \hat\mu_{c1}^2 - \frac{1}{2} \sum_{c=1}^\ell \frac{n_c}{n} \frac{S_cT_{c0}}{\#T_0} \hat\mu_{c0}^2
    \nn {}& + \frac{1}{2} \sum_{c=1}^\ell \frac{n_c}{n} \frac{S_cT_{c1}}{\#T_1}\widehat{\var}(\hat\mu_{c1}) + \frac{1}{2} \sum_{c=1}^\ell \frac{n_c}{n}  \frac{S_cT_{c0}}{\#T_0}\widehat{\var}(\hat\mu_{c0}). 
\end{align}
Taking expectation:
\begin{align}
    {}& \E\left(\widehat{\cov}_C(\hat\mu_1, \hat\mu_0)\right) =  \E\left[\E\left(\left.\sum_{c=1}^\ell \sum_{c'\neq c} \left[1  - \frac{n_cn_{c'}}{n^2}\frac{s(s-1)}{\pi_{cc'}}\right]   \frac{S_cT_{ct}S_{c'}T_{c't}}{\#T_1 \#T_0} \hat\mu_{c1}\hat\mu_{c'0} \right|\mathbf{S,T}\right)\right]
    \nn {}& \hspace{3ex} - \E\left[\E\left(\left.\frac{1}{2} \sum_{c=1}^\ell \frac{n_c}{n}\frac{S_cT_{c1}}{\#T_1}\hat\mu_{c1}^2  \right|\mathbf{S,T}\right)\right]  
     - \E\left[\E\left(\left.\frac{1}{2} \sum_{c=1}^\ell \frac{n_c}{n} \frac{S_cT_{c0}}{\#T_0} \hat\mu_{c0}^2 \right|\mathbf{S,T}\right)\right] 
    \nn {}& \hspace{3ex} + \E\left[\E\left(\left. \frac{1}{2} \sum_{c=1}^\ell \frac{n_c}{n} \frac{S_cT_{c1}}{\#T_1} \widehat{\var}(\hat\mu_{c1}) \right|\mathbf{S,T}\right)\right] 
     + \E\left[\E\left(\left. \frac{1}{2} \sum_{c=1}^\ell \frac{n_c}{n}  \frac{S_cT_{c0}}{\#T_0} \widehat{\var}(\hat\mu_{c0}) \right|\mathbf{S,T}\right)\right]
    \nn {}& = \sum_{c=1}^\ell \sum_{c'\neq c} \left[1  - \frac{n_cn_{c'}}{n^2}\frac{s(s-1)}{\pi_{cc'}}\right] \mu_{c1}\mu_{c'0} \E\left(\frac{S_cT_{ct}S_{c'}T_{c't}}{\#T_1 \#T_0} \right) 
    \nn {}& \hspace{3ex} - \frac{1}{2} \sum_{c=1}^\ell \frac{n_c}{n}\left[\mu_{c1}^2+\var(\hat\mu_{c1})\right] \E\left(\frac{S_cT_{c1}}{\#T_1} \right) - \frac{1}{2} \sum_{c=1}^\ell \frac{n_c}{n}\left[\mu_{c0}^2+\var(\hat\mu_{c1})\right] \E\left(\frac{S_cT_{c0}}{\#T_0} \right)
    \nn {}& \hspace{3ex} + \frac{1}{2} \sum_{c=1}^\ell \frac{n_c}{n}\var(\hat\mu_{c1}) \E\left(\frac{S_cT_{c1}}{\#T_1} \right) + \frac{1}{2} \sum_{c=1}^\ell \frac{n_c}{n} \var(\hat\mu_{c0}) \E\left(\frac{S_cT_{c0}}{\#T_0} \right)
    \nn {}& = \sum_{c=1}^\ell \sum_{c'\neq c} \left[\frac{\pi_{cc'}}{s(s-1)}  - \frac{n_cn_{c'}}{n^2}\right] \mu_{c1}\mu_{c'0} - \frac{1}{2} \sum_{c=1}^\ell \frac{n_c^2}{n^2} \mu_{c1}^2 - \frac{1}{2} \sum_{c=1}^\ell \frac{n_c^2}{n^2} \mu_{c0}^2. \label{covppsexp}
\end{align}
We next show that eq.~\eqref{covppsexp} is no larger than eq.~\eqref{covpps}, using Young's inequality.
\begin{lemma}[Young's Inequality]
If $a,b$ are nonnegative real numbers and $p,q$ are positive real numbers such that $\frac{1}{p} + \frac{1}{q} = 1$, then
\begin{align}
    ab \leq \frac{a^p}{p} + \frac{b^q}{q}.
\end{align}
\end{lemma}
\noindent Take $p=q=2$, then
\begin{align}
    \cov(\hat\mu_{1,\text{HT-PPS}}, \hat\mu_{0,\text{HT-PPS}}) ={}& \sum_{c=1}^\ell \sum_{c\neq c'} \left( \frac{\pi_{cc'}}{s(s-1)} - \frac{n_cn_{c'}}{n^2} \right) \mu_{c1} \mu_{c0} - \sum_{c=1}^\ell \frac{n_c^2}{n^2} \mu_{c1}\mu_{c0}
    \nn  \geq{}& \sum_{c=1}^\ell \sum_{c\neq c'} \left( \frac{\pi_{cc'}}{s(s-1)} - \frac{n_cn_{c'}}{n^2} \right) \mu_{c1} \mu_{c0}
    \nn {}&  - \frac{1}{2} \sum_{c=1}^\ell \frac{n_c^2}{n^2} \mu_{c1}^2 - \frac{1}{2} \sum_{c=1}^\ell \frac{n_c^2}{n^2} \mu_{c0}^2
    \nn ={}& \cov_{C}(\hat\mu_{1,\text{HT-PPS}}, \hat\mu_{0,\text{HT-PPS}}). \label{covbound}
\end{align}


\cleardoublepage

\section{Supplementary appendix on HT-PPS properties} \label{appendixhtppsvarlarge}

\subsection{Useful indicator properties under PPS} 
We begin this section by computing expectations, variances, and covariances of indicators under PPSWOR sampling of clusters.    Define $\pi_{cc'} \equiv E(S_cS_{c'}) = P(S_c = 1, S_{c'} = 1)$ as the probability of sampling both cluster $c$ and $c'$.
\noindent
\begin{align}
  \E(S_c|\#T_t) ={}& \frac{n_cs}{n} \\
  \E(S_c^2T_{ct}^2|\#T_t) ={}& \E(S_cT_{ct}|\#T_t) = \E(S_c \E(T_{ct}|\mathbf{S})|\#T_t) 
  \nn ={}& \E\left(\left.S_c\frac{\#T_t}{s}\right|\#T_t\right) = \frac{\#T_t}{s}\E(S_c|\#T_t) 
  \nn ={}& = \frac{n_c\#T_t}{n} \\
  \E(S_cS_{c'}T_{ct}T_{c't}|\#T_t) ={}& \E(S_cS_{c'}\E(T_{ct}T_{c't} | \mathbf{S})|\#T_t) 
  \nn ={}& \E\left(\left. S_cS_{c'}\frac{\#T_t}{s}\frac{\#T_t-1}{s-1} \right|\#T_t \right) = \frac{\#T_t(\#T_t)}{s(s-1)}\E(S_cS_{c'}|\#T_t) 
  \nn ={}&  \frac{\#T_t(\#T_t-1)}{s(s-1)}\pi_{cc'}\\
  \E(S_cS_{c'}T_{c1}T_{c'0}|\#T_1, \#T_0) ={}& \E(S_cS_{c'}\E(T_{c1}T_{c'0} | \mathbf{S})|\#T_1, \#T_0) 
  \nn ={}& \frac{\#T_1}{s}\frac{\#T_0}{s-1} \E(S_cS_{c'}|\#T_1, \#T_0) 
  \nn ={}& \frac{\#T_1 \#T_0}{s(s-1)}\pi_{cc'} \\
  \E\left(\frac{S_c^2T_{ct}^2}{\#T_t^2}\right) ={}& \E\left(\frac{1}{\#T_t^2}\E(S_cT_{ct}|\#T_t)\right) = \frac{n_c}{n}\E\left(\frac{1}{\#T_t}\right)\\
  \E\left(\frac{S_cS_{c'}T_{ct}T_{c't}}{\#T_t^2}\right) ={}& \E\left(\frac{1}{\#T_t^2}\E(S_cS_{c'}T_{ct}T_{c't}|\#T_t)\right) 
  \nn ={}& \frac{\pi_{cc'}}{s(s-1)}\E\left(1-\frac{1}{\#T_t}\right) \\
  \var(S_cT_{ct}|\#T_t) ={}& \frac{n_c\#T_t}{n}\left(1-\frac{n_c\#T_t}{n}\right)
\end{align}

Conditional on the sampled clusters, units are sampled within a cluster using simple random sampling, and this secondary sampling stage is independent of cluster treatment assignment.  Thus, the expectation of within-cluster sampling indicators are independent of the cluster treatment indicators.  Moreover, within-cluster samples are drawn independently across clusters, and so for distinct units $k$ and $k'$ in the same cluster $c$ or distinct units $k$ and $k^*$ in different clusters $c$ and $c'$:
 \begin{align} 
        \E\left( S_{kc} | \mathbf S \right) &=  \frac{s_c}{n_c}  
        \\
        \E\left( S_{kc}S_{k'c} | \mathbf S\right) &=  \frac{s_c(s_c - 1)}{n_c(n_c - 1)}   \\
        \E\left( S_{kc}S_{k^*c'} | \mathbf S\right) &= \frac{s_cs_{c'}}{n_c n_{c'}}  \\
        \var(S_{kc}| \mathbf S) &= \frac{s_c}{n_c}\left(1-\frac{s_c}{n_c}\right) \\
        \cov(S_{kc}, S_{k'c}| \mathbf S) &= \E(S_{kc}S_{k'c}| \mathbf S)- \E(S_{kc}| \mathbf S)\E(S_{k'c}| \mathbf S)
        \nn &= -\frac{s_c}{n_c}\frac{1}{n_c-1}\left(1-\frac{s_c}{n_c}\right) 
    \end{align}

\subsection{Location invariance of HT-PPS estimator for PATE}

Since,
\begin{align}
    \hat\mu_{t,\text{HT,PPS}}(a + \mathbf{y}) &= \sum_{c=1}^{\ell}  \frac{S_cT_{ct}}{\#T_t}\sum_{k=1}^{n_c} \frac{(a + y_{kct})S_{kc}}{s_c}
    \nn &= 	a\sum_{c=1}^{\ell}  \frac{S_cT_{c1}}{\#T_1}\sum_{k=1}^{n_c} \frac{S_{kc}}{s_c} + \sum_{c=1}^{\ell}  \frac{S_cT_{c1}}{\#T_1}\sum_{k=1}^{n_c} \frac{y_{kc1}S_{kc}}{s_c}
    \nn &= a + \hat\mu_{t,\text{HT,PPS}}(\mathbf{y})
\end{align}
the HT-PPS estimator for PATE is location-invariant:
\begin{equation}
    \hat\delta_\text{HT,PPS}(a+\mathbf{y}) = \hat\mu_{1,\text{HT,PPS}}(a + \mathbf{y}) - \hat\mu_{0,\text{HT,PPS}}(a + \mathbf{y}) = \hat\delta_\text{HT,PPS}(\mathbf{y}).
\end{equation}
   
\subsection{Expectation of HT-PPS estimator for PATE}

  \begin{align}
  \E(\hat\delta_{\text{HT-PPS}}) ={}& \E\left(\sum_{c=1}^{\ell}  \frac{S_cT_{c1}}{\#T_1}\sum_{k=1}^{n_c} \frac{y_{kc1}S_{kc}}{s_c} - \sum_{c=1}^{\ell}  \frac{S_{c'}T_{c'0}}{\#T_0}\sum_{k^*=1}^{n_{c'}} \frac{y_{k^*c'0}S_{k^*c'}}{s_{c'}}\right) 
  \nn ={}& \sum_{c=1}^{\ell} \sum_{k=1}^{n_c} \frac{y_{kc1}}{s_c}\E\left( \frac{S_cT_{c1}S_{kc}}{\#T_1}  \right) - \sum_{c'=1}^{\ell} \sum_{k^*=1}^{n_{c'}} \frac{y_{k^*c'0}}{s_{c'}}\E\left( \frac{S_{c'}T_{c'0}S_{k^*c'}}{\#T_0}  \right) 
    \nn ={}& \sum_{c=1}^{\ell} \sum_{k=1}^{n_c} \frac{y_{kc1}}{ s_c}\E\left( S_c\E\left(\left.\frac{T_{c1}}{\#T_1}\right|\mathbf{S}\right)\E(S_{kc}|\mathbf{S})  \right) 
    \nn {}& - \sum_{c'=1}^{\ell} \sum_{k^*=1}^{n_{c'}} \frac{y_{k^*c'0}}{ s_{c'}}\E\left( S_{c'}\E\left(\left.\frac{T_{c'0}}{\#T_0}\right|\mathbf{S}\right)\E(S_{k^*c'}|\mathbf{S})  \right)
    \nn ={}& \sum_{c=1}^{\ell} \sum_{k=1}^{n_c} \frac{y_{kc1}}{n_cs}\E\left(S_c\right)  - \sum_{c'=1}^{\ell} \sum_{k^*=1}^{n_{c'}} \frac{y_{k^*c'0}}{n_{c'}s}\E\left(S_{c'}\right) 
    \nn ={}& \sum_{c=1}^{\ell} \frac{n_c}{n}\mu_{c1} - \sum_{c'=1}^{\ell} \frac{n_{c'}}{n}\mu_{c'0} 
    \nn ={}& \mu_1-\mu_0=\delta.
  \end{align}

\subsection{Variance of HT-PPS estimator for PATE}

From the property
\begin{equation}
    \var(\hat\delta) = \var(\hat\mu_1 - \hat\mu_0) = \var(\hat\mu_1) + \var(\hat\mu_0) - 2\cov(\hat\mu_1, \hat\mu_0).
\end{equation}
each term is expanded upon to derive the variance of the HT-PPS estimator for PATE and obtain a variance estimator.

\subsubsection{Variance of HT-PPS estimator for population mean} 

Using the law of total variance,
   \begin{align} 
       \var(\hat\mu_{t,\text{HT,PPS}}) ={}& \var\left(\sum_{c=1}^\ell \frac{S_cT_{ct}}{\#T_t} \sum_{k=1}^{n_c}\frac{y_{kct}S_{kc}}{s_c} \right)
       \nn ={}& \var\left[ \E\left(\left.\sum_{c=1}^\ell \frac{S_cT_{ct}}{\#T_t}\sum_{k=1}^{n_c}\frac{y_{kct}S_{kc}}{s_c} \right| \mathbf{S, T} \right) \right] 
       \nn {}& + \E\left[\var\left(\left.\sum_{c=1}^\ell \frac{S_cT_{ct}}{\#T_t}\sum_{k=1}^{n_c}\frac{y_{kct}S_{kc}}{s_c} \right| \mathbf{S, T} \right) \right].
\end{align}
The first terms can be further simplified:
\begin{align}
       {}&  \var\left[ \E\left(\left.\sum_{c=1}^\ell \frac{S_cT_{ct}}{\#T_t}\sum_{k=1}^{n_c}\frac{y_{kct}S_{kc}}{s_c} \right| \mathbf{S, T} \right) \right] = \var\left[\sum_{c=1}^\ell \frac{S_cT_{ct}}{\#T_t} \sum_{k=1}^{n_c}\frac{y_{kct}}{s_c} \E(S_{kc}|\mathbf{S,T}) \right]
       \nn {}& = \sum_{c=1}^\ell \var\left(\mu_{ct} \frac{S_cT_{ct}}{\#T_t}\right) + \sum_{c=1}^\ell \sum_{c'\neq c} \cov\left(\mu_{ct} \frac{S_cT_{ct}}{\#T_t}, \mu_{c't} \frac{S_{c'}T_{c't}}{\#T_t}\right)
       \nn {}& = \sum_{c=1}^\ell \mu_{ct}^2 \var\left(\frac{S_cT_{ct}}{\#T_t}\right)  + \sum_{c=1}^\ell \sum_{c'\neq c} \mu_{ct}\mu_{c't}\cov\left(\frac{S_cT_{ct}}{\#T_t}, \frac{S_{c'}T_{c't}}{\#T_t}\right)
       \nn {}&= \sum_{c=1}^\ell \mu_{ct}^2 \left[\E\left(\frac{S_cT_{ct}}{\#T_t}\right) - \E\left(\frac{S_cT_{ct}}{\#T_t}\right)^2 \right] 
       \nn {}& \hspace{3ex} + \sum_{c=1}^\ell \sum_{c'\neq c} \mu_{ct}\mu_{c't} \left[\E\left(\frac{S_cS_{c'}T_{ct}T_{c't}}{\#T_t^2}\right)-\E\left(\frac{S_cT_{ct}}{\#T_t}\right)\E\left(\frac{S_{c'}T_{c't}}{\#T_t}\right) \right]
       \nn {}& = \E\left(\frac{1}{\#T_t}\right) \sum_{c=1}^\ell \frac{n_c}{n} \mu_{ct}^2 -\sum_{c=1}^\ell \sum_{c'\neq c} \frac{n_c^2}{n^2}\mu_{ct}^2 
       \nn {}& \hspace{3ex} + \E\left(1-\frac{1}{\#T_t}\right)\sum_{c=1}^\ell \sum_{c'\neq c}\frac{\pi_{cc'}}{s(s-1)} \mu_{ct}\mu_{c't} - \sum_{c=1}^\ell \sum_{c'\neq c}\frac{n_c^2}{n^2}\mu_{ct}\mu_{c't}
       \nn {}& = \E\left(\frac{1}{\#T_t}\right) \sum_{c=1}^\ell \frac{n_c}{n} \mu_{ct}^2  + \E\left(1-\frac{1}{\#T_t}\right)\sum_{c=1}^\ell \sum_{c'\neq c}\frac{\pi_{cc'}}{s(s-1)} \mu_{ct}\mu_{c't} - \mu_t^2
       \nn {}& = \E\left(\frac{1}{\#T_t}\right) \sum_{c=1}^\ell \frac{n_c}{n} \mu_{ct}^2 - \E\left(\frac{1}{\#T_t}\right)\mu_t^2  
       \nn {}& \hspace{3ex} + \E\left(1-\frac{1}{\#T_t}\right)\sum_{c=1}^\ell \sum_{c'\neq c}\frac{\pi_{cc'}}{s(s-1)} \mu_{ct}\mu_{c't} - \E\left(1-\frac{1}{\#T_t}\right) \mu_t^2
       \nn {}& = \E\left(\frac{1}{\#T_t}\right)\left[\sum_{c=1}^\ell \frac{n_c}{n}\mu_{ct}^2 - \mu_t^2 \right] + \E\left(1-\frac{1}{\#T_t}\right) \left[\sum_{c=1}^\ell \sum_{c'\neq c}\frac{\pi_{cc'}}{s(s-1)}\mu_{ct}\mu_{c't} - \mu_t^2 \right]
       \nn {}& = \E\left(\frac{1}{\#T_t}\right)\left[\sum_{c=1}^\ell \frac{n_c}{n}(\mu_{ct}^2 -2\mu_t\mu_{ct} + \mu_t^2) \right] + \E\left(1-\frac{1}{\#T_t}\right) \left[\sum_{c=1}^\ell \sum_{c'\neq c}\frac{\pi_{cc'}}{s(s-1)}\mu_{ct}\mu_{c't} - \mu_t^2 \right]
       \nn {}& = \E\left(\frac{1}{\#T_t}\right)\left[\sum_{c=1}^\ell \frac{n_c}{n}(\mu_{ct} - \mu_t)^2 \right] + \E\left(1-\frac{1}{\#T_t}\right) \left[\sum_{c=1}^\ell \sum_{c'\neq c}\frac{\pi_{cc'}}{s(s-1)}\mu_{ct}\mu_{c't} - \mu_t^2 \right]
       \nn {}& = \E\left(\frac{1}{\#T_t}\right)\sigma^2_{t,bet} + \E\left(1-\frac{1}{\#T_t}\right) \left[\sum_{c=1}^\ell \sum_{c'\neq c}\frac{\pi_{cc'}}{s(s-1)}\mu_{ct}\mu_{c't} - \mu_t^2 \right]
\end{align}   
where $\sigma^2_{t,bet}$ is the weighted variance of cluster means.  Simplifying the second term:
\begin{align}       
       & \E\left[\var\left(\left.\sum_{c=1}^\ell \frac{S_cT_{ct}}{\#T_t}\sum_{k=1}^{n_c}\frac{y_{kct}S_{kc}}{s_c} \right| \mathbf{S, T} \right) \right] =  \sum_{c=1}^\ell \var(\hat\mu_{ct}|\mathbf{S, T}) \E\left(\frac{S_cT_{ct}}{\#T_t^2}\right)
       \nn {}& = \E\left(\frac{1}{\#T_t}\right) \sum_{c=1}^\ell \frac{n_c}{n}\var(\hat\mu_{ct})
        \nn {}& =  \E\left(\frac{1}{\#T_t}\right) \sum_{c=1}^\ell \frac{n_c}{n} \left[ \var\left( \sum_{k=1}^{n_c} \frac{y_{kct}S_{kc}}{s_c} \right)  + \cov\left( \sum_{k=1}^{n_c} \frac{y_{kct}S_{kc}}{s_c}, \sum_{k'\neq k} \frac{y_{k'ct}S_{k'c}}{s_c} \right) \right] 
        \nn {}& = \E\left(\frac{1}{\#T_t}\right) \sum_{c=1}^\ell \frac{n_c}{n} \left[ \sum_{k=1}^{n_c} \frac{y_{kct}^2}{s_cn_c}\left(1-\frac{s_c}{n_c} \right) - \sum_{k=1}^{n_c}\sum_{k'\neq k}\frac{y_{kct}y_{k'ct}}{s_cn_c(n_c-1)}\left(1-\frac{s_c}{n_c}\right) \right]
       \nn {}& = \E\left( \frac{1}{\#T_t}\right) \sum_{c=1}^\ell \frac{1}{n(n_c-1)s_c}\left(1-\frac{s_c}{n_c}\right)\left[(n_c-1)\sum_{k=1}^{n_c}y_{kct}^2 - \sum_{c=1}^{n_c}\sum_{c\neq c'}y_{kct}y_{kc't} \right]
       \nn {}& = \E\left( \frac{1}{\#T_t}\right) \sum_{c=1}^\ell \frac{1}{n(n_c-1)s_c}\left(1-\frac{s_c}{n_c}\right) \left[ (n_c-1)\sum_{k=1}^{n_c}y_{kct}^2-\sum_{k=1}^{n_c}\sum_{k'=1}^{n_c}y_{kct}y_{k'ct} + \sum_{k=1}^{n_c}y_{kct}^2 \right]
       \nn {}& = \E\left( \frac{1}{\#T_t} \right) \sum_{c=1}^\ell \frac{1}{n(n_c-1)s_c}\left(1-\frac{s_c}{n_c}\right) \left[ n_c\sum_{k=1}^{n_c}y_{kct}^2-\left(\sum_{k=1}^{n_c}y_{kct}\right)^2 \right]
       \nn {}& = \E\left( \frac{1}{\#T_t}\right) \sum_{c=1}^\ell \frac{n_c}{n}\left(1-\frac{s_c}{n_c}\right) \frac{\sigma^2_{ct}}{s_c}.
       \end{align}
The variance for the HT-PPS mean estimator is
\begin{align}
    \var(\hat\mu_{t, \text{HT, SRS}}) ={}& \E\left(\frac{1}{\#T_t}\right) \sum_{c=1}^\ell \frac{n_c}{n}\mu_{ct}^2  + \E\left(1-\frac{1}{\#T_t}\right)\sum_{c=1}^\ell \sum_{c\neq c'} \frac{\pi_{cc'}}{s(s-1)}  \mu_{ct}\mu_{c't} - \mu_t^2 
    \nn {}& + \E\left(\frac{1}{\#T_t}\right)\sum_{c=1}^\ell \frac{n_c}{n}\var(\hat\mu_{ct}) \label{varpps2} 
    \\
     ={}& \E\left(\frac{1}{\#T_t}\right)\sigma^2_{t,bet} + \E\left(1-\frac{1}{\#T_t}\right) \left[\sum_{c=1}^\ell \sum_{c'\neq c}\frac{\pi_{cc'}}{s(s-1)}\mu_{ct}\mu_{c't} - \mu_t^2 \right] 
    \nn {}& + \E\left( \frac{1}{\#T_t}\right) \sum_{c=1}^\ell \frac{n_c}{n}\left(1-\frac{s_c}{n_c}\right) \frac{\sigma^2_{ct}}{s_c}.
\end{align}

\subsubsection{Covariance of HT-PPS estimator for population means}	
	Note that:
   \begin{align}
       \hat \mu_{1,\text{HT,PPS}}\hat \mu_{0,\text{HT,PPS}} & = \left( \sum_{c=1}^{\ell} \frac{S_cT_{c1}}{\#T_1}\sum_{k=1}^{n_c} \frac{y_{kc1}S_{kc}}{ s_c}\right) \left( \sum_{c'=1}^{\ell} \frac{S_{c'}T_{c'0}}{\#T_0}\sum_{k^*=1}^{n_{c'}} \frac{y_{k*c'0}S_{k*c'}}{ s_{c'}}\right)
       \nn &= \sum_{c=1}^{\ell} \sum_{k=1}^{n_c}\sum_{c'=1}\sum_{k^*=1}^{n_{c'}} \frac{y_{kc1}y_{k^*c'0}}{s_cs_{c'}} \frac{S_cT_{c1}S_{kc}S_{c'}T_{c'0}S_{k^*c'}}{\#T_1\#T_0}
        \nn &= \sum_{c=1}^{\ell} \sum_{k=1}^{n_c}\sum_{c'\neq c} \sum_{k^*=1}^{n_{c'}} \frac{y_{kc1}y_{k^*c'0}}{s_cs_{c'}} \frac{S_cT_{c1}S_{kc}S_{c'}T_{c'0}S_{k^*c'}}{\#T_1\#T_0}.
   \end{align}
The last equality comes from the fact that a cluster can only be given one treatment.   
Therefore,
   \begin{align}
       {}& \cov(\hat \mu_{1,\text{HT,PPS}}, \hat \mu_{0,\text{HT,PPS}})  = \E(\hat \mu_{1,\text{HT,PPS}}\hat \mu_{0,\text{HT,PPS}}) - \E(\hat \mu_{1,\text{HT,PPS}})\E(\hat \mu_{0,\text{HT,PPS}})
       \nn {}& = \E\left(\sum_{c=1}^{\ell} \sum_{k=1}^{n_c}\sum_{c'\neq c} \sum_{k^*=1}^{n_{c'}} \frac{y_{kc1}y_{k^*c'0}}{s_cs_{c'}} \frac{S_cT_{c1}S_{kc}S_{c'}T_{c'0}S_{k^*c'}}{\#T_1\#T_0}\right) - \mu_1\mu_0
       \nn {}& = \sum_{c=1}^{\ell} \sum_{k=1}^{n_c}\sum_{c'\neq c} \sum_{k^*=1}^{n_{c'}} \frac{y_{kc1}y_{k^*c'0}}{s_cs_{c'}} \E\left[\E\left(\frac{S_cT_{c1}S_{kc}S_{c'}T_{c'0}S_{k^*c'}}{\#T_1\#T_0}|\mathbf{S}\right)\right] - \mu_1\mu_0
       \nn {}&= \sum_{c=1}^{\ell} \sum_{k=1}^{n_c}\sum_{c'\neq c} \sum_{k^*=1}^{n_{c'}} \frac{y_{kc1}y_{k^*c'0}}{s_cs_{c'}}  \E\left[S_cS_{c'} \E\left(\left.\frac{T_{c1}T_{c'0}}{\#T_1\#T_0}\right|\mathbf{S}\right) \E(S_{kc}S_{k^*c'}|\mathbf{S})\right] - \mu_1\mu_0
       \nn {}&= \frac{1}{s(s-1)} \sum_{c=1}^{\ell} \sum_{k=1}^{n_c}\sum_{c'\neq c} \sum_{k^*=1}^{n_{c'}} \frac{y_{kc1}y_{k^*c'0}}{n_cn_{c'}}  \E\left(S_cS_{c'}\right) - \mu_1\mu_0
       \nn {}&= \frac{1}{s(s-1)}\sum_{c=1}^{\ell} \sum_{c'\neq c}  \pi_{cc'}\mu_{c1}\mu_{c'0} - \mu_1\mu_0
       \nn {}&= \sum_{c=1}^{\ell} \sum_{c'\neq c}  \left[\frac{\pi_{cc'}}{s(s-1)} - \frac{n_cn_{c'}}{n^2}\right] \mu_{c1}\mu_{c'0} - \sum_{c=1}^\ell \frac{n_c^2}{n^2} \mu_{c1}\mu_{c0}. \label{covpps2}
   \end{align}

\subsubsection{SYG estimator for variance}

The SYG variance estimator is
\begin{align}
    \widehat\var(\hat\mu_t) ={}& \frac{1}{2} \sum_{c=1}^\ell \sum_{c\neq c'} \left[\frac{s(s-1)}{\pi_{cc'}\#T_t(\#T_t-1)}\frac{n_cn_{c'}}{n^2} - \frac{1}{\#T_t^2} \right] S_cT_{ct}S_{c'}T_{c't} (\hat\mu_{ct}-\hat\mu_{c't})^2 
    \nn {}& +  \sum_{c=1}^\ell \frac{S_cT_{ct}}{\#T_t} \frac{n_c}{n}  \widehat{\var}(\hat\mu_{ct})
\end{align}
where 
\begin{equation}
    \widehat{\var}(\hat\mu_{ct}) = \left(1-\frac{s_c}{n_c}\right) \frac{\hat\sigma^2_{ct}}{s_c}.
\end{equation}
The $\hat\sigma^2_{ct}$ is the sample variance of outcomes, which is unbiased for the population variance $\sigma^2_{ct}$.  We will now show that the SYG variance is unbiased for $\var(\hat\mu_t)$.  This requires the following:
\begin{equation}
    \sum_{c'\neq c}^\ell n_{c'} = n-n_c
\end{equation}
and
\begin{equation}
    \sum_{c'\neq c}^\ell \pi_{cc'} = \sum_{c'\neq c}^\ell E(S_cS_{c'}) = E[S_c(s-S_c)] = \frac{n_cs}{n}(s-1).
\end{equation}
Therefore, the expectation is
\begin{align}
    {}& \E\left(\widehat\var(\hat\mu_t)\right) = \E\left( \frac{1}{2} \sum_{c=1}^\ell \sum_{c\neq c'} \left[ \frac{s(s-1)}{\pi_{cc'}} \frac{n_cn_{c'}}{n^2} \frac{S_cS_{c'}T_{ct}T_{c't}}{\#T_t(\#T_t-1)} - \frac{S_cS_{c'}T_{ct}T_{c't}}{\#T_t^2} \right]  [\hat\mu_{ct}-\hat\mu_{c't}]^2 \right.
    \nn {}& \hspace{3ex} \left. +  \sum_{c=1}^\ell \frac{n_c}{n} \frac{S_cT_{ct}}{\#T_t}\widehat{\var}(\hat\mu_{ct})\right)
    \nn {}& = \E\left( \E\left(\left.\frac{1}{2} \sum_{c=1}^\ell \sum_{c\neq c'} \left[ \frac{s(s-1)}{\pi_{cc'}} \frac{n_cn_{c'}}{n^2} \frac{S_cS_{c'}T_{ct}T_{c't}}{\#T_t(\#T_t-1)} - \frac{S_cS_{c'}T_{ct}T_{c't}}{\#T_t^2} \right]  [\hat\mu_{ct}-\hat\mu_{c't}]^2\right|\mathbf{S, T}\right)\right)
    \nn {}& \hspace{3ex} + \E\left(\E\left(\left. \sum_{c=1}^\ell \frac{n_c}{n} \frac{S_cT_{ct}}{\#T_t}\widehat{\var}(\hat\mu_{ct})\right|\mathbf{S,T}\right)\right)
    \nn {}& = \E\left( \E\left(\left. \sum_{c=1}^\ell \sum_{c\neq c'} \left[ \frac{s(s-1)}{\pi_{cc'}} \frac{n_cn_{c'}}{n^2} \frac{S_cS_{c'}T_{ct}T_{c't}}{\#T_t(\#T_t-1)} - \frac{S_cS_{c'}T_{ct}T_{c't}}{\#T_t^2} \right]  [\hat\mu^2_{ct}-\hat\mu_{ct}\hat\mu_{c't}]\right|\mathbf{S, T}\right)\right)
    \nn {}& \hspace{3ex} + \E\left(\sum_{c=1}^\ell \frac{n_c}{n} \frac{S_cT_{ct}}{\#T_t}\var(\hat\mu_{ct})\right)
    \nn ={}& \E\left( \sum_{c=1}^\ell \sum_{c\neq c'} \left[ \frac{s(s-1)}{\pi_{cc'}} \frac{n_cn_{c'}}{n^2} \frac{S_cS_{c'}T_{ct}T_{c't}}{\#T_t(\#T_t-1)} - \frac{S_cS_{c'}T_{ct}T_{c't}}{\#T_t^2} \right]  [\mu^2_{ct}+\var(\hat\mu_{ct})-\mu_{ct}\mu_{c't}]\right)
    \nn {}& + \E\left( \sum_{c=1}^\ell \frac{n_c}{n} \frac{S_cT_{ct}}{\#T_t}\var(\hat\mu_{ct})\right) 
    \nn ={}&  \sum_{c=1}^\ell \sum_{c\neq c'} \left[ \frac{s(s-1)}{\pi_{cc'}} \frac{n_cn_{c'}}{n^2} \E\left(\frac{S_cS_{c'}T_{ct}T_{c't}}{\#T_t(\#T_t-1)}\right) - \E\left(\frac{S_cS_{c'}T_{ct}T_{c't}}{\#T_t^2}\right) \right]  [\mu^2_{ct}+\var(\hat\mu_{ct})-\mu_{ct}\mu_{c't}]
    \nn {}& +  \sum_{c=1}^\ell \frac{n_c}{n} \var(\hat\mu_{ct}) \E\left(\frac{S_cT_{ct}}{\#T_t}\right) 
    \nn ={}&  \sum_{c=1}^\ell \sum_{c\neq c'} \left[ \frac{s(s-1)}{\pi_{cc'}} \frac{n_cn_{c'}}{n^2} \E\left(\frac{1}{\#T_t(\#T_t-1)}\E\left(\left.S_cS_{c'}T_{ct}T_{c't} \right| \#T_t \right)\right) - \E\left( \frac{1}{\#T_t^2}\E\left(\left.S_cS_{c'}T_{ct}T_{c't} \right| \#T_t\right)\right) \right]  
    \nn {}& \hspace{3ex} \cdot  [\mu^2_{ct}+\var(\hat\mu_{ct})-\mu_{ct}\mu_{c't}] +  \sum_{c=1}^\ell \frac{n_c}{n} \var(\hat\mu_{ct}) \E\left(\frac{1}{\#T_t} \E\left(S_cT_{ct}|\#T_t\right) \right)
    \nn ={}& \sum_{c=1}^\ell \sum_{c\neq c'} \left[\frac{n_cn_{c'}}{n^2} - \frac{\pi_{cc'}}{s(s-1)}\E\left(1-\frac{1}{\#T_t}\right) \right]  (\mu^2_{ct}+\var(\hat\mu_{ct})-\mu_{ct}\mu_{c't}) + \sum_{c=1}^\ell \frac{n_c^2}{n^2}\var(\hat\mu_{ct})  
    \nn ={}& \sum_{c=1}^\ell \sum_{c'\neq c} \left[\frac{n_cn_{c'}}{n^2} - \frac{\pi_{cc'}}{s(s-1)}\E\left(1-\frac{1}{\#T_t}\right) \right]  (\mu^2_{ct}+\var(\hat\mu_{ct})) 
    \nn {}& - \sum_{c=1}^\ell \sum_{c\neq c'} \left[\frac{n_cn_{c'}}{n^2} - \frac{\pi_{cc'}}{s(s-1)}\E\left(1-\frac{1}{\#T_t}\right) \right] \mu_{ct}\mu_{c't} + \sum_{c=1}^\ell \frac{n_c^2}{n^2}\var(\hat\mu_{ct})
    \nn ={}& \sum_{c=1}^\ell \left[\frac{n_c}{n^2}\sum_{c'\neq c} n_{c'} - \frac{1}{s(s-1)}\E\left(1-\frac{1}{\#T_t}\right)\sum_{c'\neq c} \pi_{cc'}\right] (\mu^2_{ct}+\var(\hat\mu_{ct})) 
    \nn {}& - \sum_{c=1}^\ell \sum_{c\neq c'} \left[\frac{n_cn_{c'}}{n^2} - \frac{\pi_{cc'}}{s(s-1)}\E\left(1-\frac{1}{\#T_t}\right) \right] \mu_{ct}\mu_{c't} + \sum_{c=1}^\ell \frac{n_c^2}{n^2}\var(\hat\mu_{ct})
    \nn ={}& \sum_{c=1}^\ell \left[\frac{n_c}{n}\left(1-\frac{n_c}{n}\right) - \frac{n_c}{n}\E\left(1-\frac{1}{\#T_t}\right)\right] (\mu^2_{ct}+\var(\hat\mu_{ct}))
    \nn {}& - \sum_{c=1}^\ell \sum_{c\neq c'} \left[\frac{n_cn_{c'}}{n^2} - \frac{\pi_{cc'}}{s(s-1)}\E\left(1-\frac{1}{\#T_t}\right) \right] \mu_{ct}\mu_{c't} + \sum_{c=1}^\ell \frac{n_c^2}{n^2}\var(\hat\mu_{ct})
    \nn ={}& \E\left(\frac{1}{\#T_t}\right) \sum_{c=1}^\ell \frac{n_c}{n}\mu_{ct}^2 - \sum_{c=1}^\ell \frac{n_c^2}{n^2}\mu_{ct}^2 - \sum_{c=1}^\ell \sum_{c\neq c'} \frac{n_cn_{c'}}{n^2}\mu_{ct}\mu_{c't} 
    \nn {}& + \E\left(1-\frac{1}{\#T_t}\right)\sum_{c=1}^\ell \sum_{c\neq c'} \frac{\pi_{cc'}}{s(s-1)}  \mu_{ct}\mu_{c't} + \E\left(\frac{1}{\#T_t}\right)\sum_{c=1}^\ell \frac{n_c}{n}\var(\hat\mu_{ct})
    \nn ={}& \E\left(\frac{1}{\#T_t}\right) \sum_{c=1}^\ell \frac{n_c}{n}\mu_{ct}^2  + \E\left(1-\frac{1}{\#T_t}\right)\sum_{c=1}^\ell \sum_{c\neq c'} \frac{\pi_{cc'}}{s(s-1)}  \mu_{ct}\mu_{c't} - \mu_t^2 
    \nn {}& + \E\left(\frac{1}{\#T_t}\right)\sum_{c=1}^\ell \frac{n_c}{n}\var(\hat\mu_{ct}). \label{varppsexp2}
\end{align}
This is equal to eq.~\eqref{varpps2}.
 
\subsubsection{Covariance bound} 

The covariance is bounded by
\begin{align}
    \widehat{\cov}_C(\hat\mu_1, \hat\mu_0) ={}& \sum_{c=1}^\ell \sum_{c'\neq c} \left[1 -\frac{n_cn_{c'}}{n^2}\frac{s(s-1)}{\pi_{cc'}}\right]   \frac{S_cT_{ct}S_{c'}T_{c't}}{\#T_1 \#T_0} \hat\mu_{c1}\hat\mu_{c'0}
    \nn {}& - \frac{1}{2} \sum_{c=1}^\ell \frac{n_c}{n} \frac{S_cT_{c1}}{\#T_1} \hat\mu_{c1}^2 - \frac{1}{2} \sum_{c=1}^\ell \frac{n_c}{n} \frac{S_cT_{c0}}{\#T_0} \hat\mu_{c0}^2
    \nn {}& + \frac{1}{2} \sum_{c=1}^\ell \frac{n_c}{n} \frac{S_cT_{c1}}{\#T_1}\widehat{\var}(\hat\mu_{c1}) + \frac{1}{2} \sum_{c=1}^\ell \frac{n_c}{n}  \frac{S_cT_{c0}}{\#T_0}\widehat{\var}(\hat\mu_{c0}). 
\end{align}
Taking expectation:
\begin{align}
    {}& \E\left(\widehat{\cov}_C(\hat\mu_1, \hat\mu_0)\right) =  \E\left[\E\left(\left.\sum_{c=1}^\ell \sum_{c'\neq c} \left[1  - \frac{n_cn_{c'}}{n^2}\frac{s(s-1)}{\pi_{cc'}}\right]   \frac{S_cT_{ct}S_{c'}T_{c't}}{\#T_1 \#T_0} \hat\mu_{c1}\hat\mu_{c'0} \right|\mathbf{S,T}\right)\right]
    \nn {}& \hspace{3ex} - \E\left[\E\left(\left.\frac{1}{2} \sum_{c=1}^\ell \frac{n_c}{n}\frac{S_cT_{c1}}{\#T_1}\hat\mu_{c1}^2  \right|\mathbf{S,T}\right)\right]  
     - \E\left[\E\left(\left.\frac{1}{2} \sum_{c=1}^\ell \frac{n_c}{n} \frac{S_cT_{c0}}{\#T_0} \hat\mu_{c0}^2 \right|\mathbf{S,T}\right)\right] 
    \nn {}& \hspace{3ex} + \E\left[\E\left(\left. \frac{1}{2} \sum_{c=1}^\ell \frac{n_c}{n} \frac{S_cT_{c1}}{\#T_1} \widehat{\var}(\hat\mu_{c1}) \right|\mathbf{S,T}\right)\right] 
     + \E\left[\E\left(\left. \frac{1}{2} \sum_{c=1}^\ell \frac{n_c}{n}  \frac{S_cT_{c0}}{\#T_0} \widehat{\var}(\hat\mu_{c0}) \right|\mathbf{S,T}\right)\right]
    \nn {}& = \sum_{c=1}^\ell \sum_{c'\neq c} \left[1  - \frac{n_cn_{c'}}{n^2}\frac{s(s-1)}{\pi_{cc'}}\right] \mu_{c1}\mu_{c'0} \E\left(\frac{S_cT_{ct}S_{c'}T_{c't}}{\#T_1 \#T_0} \right) 
    \nn {}& \hspace{3ex} - \frac{1}{2} \sum_{c=1}^\ell \frac{n_c}{n}\left[\mu_{c1}^2+\var(\hat\mu_{c1})\right] \E\left(\frac{S_cT_{c1}}{\#T_1} \right) - \frac{1}{2} \sum_{c=1}^\ell \frac{n_c}{n}\left[\mu_{c0}^2+\var(\hat\mu_{c1})\right] \E\left(\frac{S_cT_{c0}}{\#T_0} \right)
    \nn {}& \hspace{3ex} + \frac{1}{2} \sum_{c=1}^\ell \frac{n_c}{n}\var(\hat\mu_{c1}) \E\left(\frac{S_cT_{c1}}{\#T_1} \right) + \frac{1}{2} \sum_{c=1}^\ell \frac{n_c}{n} \var(\hat\mu_{c0}) \E\left(\frac{S_cT_{c0}}{\#T_0} \right)
    \nn {}& = \sum_{c=1}^\ell \sum_{c'\neq c} \left[1  - \frac{n_cn_{c'}}{n^2}\frac{s(s-1)}{\pi_{cc'}}\right] \mu_{c1}\mu_{c'0} \E\left[\frac{1}{\#T_1 \#T_0} \E\left(\left.S_cS_{c'}T_{ct}T_{c't} \right|\#T_1, \#T_0 \right) \right]
    \nn {}& \hspace{3ex} - \frac{1}{2} \sum_{c=1}^\ell \frac{n_c}{n}\left[\mu_{c1}^2+\var(\hat\mu_{c1})\right] \E\left[\frac{1}{\#T_1}\E\left(S_cT_{c1}|\#T_1 \right)\right] 
    \nn {}& \hspace{3ex}- \frac{1}{2} \sum_{c=1}^\ell \frac{n_c}{n}\left[\mu_{c0}^2+\var(\hat\mu_{c1})\right] \E\left[\frac{1}{\#T_0}\E\left(S_cT_{c0}|\#T_0 \right)\right]
    \nn {}& \hspace{3ex} + \frac{1}{2} \sum_{c=1}^\ell \frac{n_c}{n}\var(\hat\mu_{c1}) \E\left[\frac{1}{\#T_1}\E\left(S_cT_{c1}|\#T_1 \right)\right] + \frac{1}{2} \sum_{c=1}^\ell \frac{n_c}{n} \var(\hat\mu_{c0}) \E\left[\frac{1}{\#T_0}\E\left(S_cT_{c0}|\#T_0 \right)\right]
    \nn {}& = \sum_{c=1}^\ell \sum_{c'\neq c} \left[\frac{\pi_{cc'}}{s(s-1)}  - \frac{n_cn_{c'}}{n^2}\right] \mu_{c1}\mu_{c'0} - \frac{1}{2} \sum_{c=1}^\ell \frac{n_c^2}{n^2} \mu_{c1}^2 - \frac{1}{2} \sum_{c=1}^\ell \frac{n_c^2}{n^2} \mu_{c0}^2. \label{covppsexp2}
\end{align}
We next show that eq.~\eqref{covppsexp2} is no larger than eq.~\eqref{covpps2}, using Young's inequality.
\begin{lemma}[Young's Inequality]
If $a,b$ are nonnegative real numbers and $p,q$ are positive real numbers such that $\frac{1}{p} + \frac{1}{q} = 1$, then
\begin{align}
    ab \leq \frac{a^p}{p} + \frac{b^q}{q}.
\end{align}
\end{lemma}
\noindent Take $p=q=2$, then
\begin{align}
    \cov(\hat\mu_{1,\text{HT-PPS}}, \hat\mu_{0,\text{HT-PPS}}) ={}& \sum_{c=1}^\ell \sum_{c\neq c'} \left( \frac{\pi_{cc'}}{s(s-1)} - \frac{n_cn_{c'}}{n^2} \right) \mu_{c1} \mu_{c0} - \sum_{c=1}^\ell \frac{n_c^2}{n^2} \mu_{c1}\mu_{c0}
    \nn  \geq{}& \sum_{c=1}^\ell \sum_{c\neq c'} \left( \frac{\pi_{cc'}}{s(s-1)} - \frac{n_cn_{c'}}{n^2} \right) \mu_{c1} \mu_{c0}
    \nn {}&  - \frac{1}{2} \sum_{c=1}^\ell \frac{n_c^2}{n^2} \mu_{c1}^2 - \frac{1}{2} \sum_{c=1}^\ell \frac{n_c^2}{n^2} \mu_{c0}^2
    \nn ={}& \cov_{C}(\hat\mu_{1,\text{HT-PPS}}, \hat\mu_{0,\text{HT-PPS}}). \label{covbound2}
\end{align}

From eq.~\eqref{varppsexp2} and eq.~\eqref{covbound2}, we see that 
\begin{align}
    \widehat{\var}(\hat\delta_\text{HT,PPS}) ={}& \frac{1}{2} \sum_{c=1}^\ell \sum_{c' \neq c} \left[\frac{s(s-1)}{\pi_{cc'}\#T_1(\#T_1-1)}\frac{n_cn_{c'}}{n^2} - \frac{1}{\#T_1^2} \right] S_cT_{c1}S_{c'}T_{c'1} \left(\hat\mu_{c1} - \hat\mu_{c'1}\right)^2 
    \nn {}& + \frac{1}{2} \sum_{c=1}^\ell \sum_{c' \neq c} \left[\frac{s(s-1)}{\pi_{cc'}\#T_0(\#T_0-1)}\frac{n_cn_{c'}}{n^2} - \frac{1}{\#T_0^2} \right] S_cT_{c0}S_{c'}T_{c'0} \left(\hat\mu_{c0} - \hat\mu_{c'0}\right)^2 
    \nn {}& - 2\sum_{c=1}^\ell \sum_{c\neq c'} \left[ \frac{\pi_{cc'}}{s(s-1)} - \frac{n_cn_{c'}}{n^2} \right] \frac{s(s-1)}{\pi_{cc'}} \frac{S_c S_{c'} T_{c1}T_{c'0}}{\#T_1\#T_0} \hat\mu_{c1} \hat\mu_{c0}  
    \nn {}& +  \sum_{c=1}^\ell \frac{n_c}{n} \frac{S_cT_{c1}}{\#T_1}\hat\mu_{c1}^2 +  \sum_{c=1}^\ell \frac{n_c}{n} \frac{S_cT_{c0}}{\#T_0} \hat\mu_{c0}^2
\end{align}
is a conservative bound for $\var(\hat\delta_\text{HT,PPS})$.

\end{document}